\documentclass[twocolumn]{aastex7}
\usepackage{amsmath}
\usepackage{booktabs}
\usepackage{makecell}

\usepackage[utf8]{inputenc}
\usepackage{verbatim}

\shorttitle{Design for Dynamic Fitness}
\shortauthors{Garcia et al.}

\begin{document}

\title{Design for Dynamic Fitness: Archetypes of urban water systems }

\author[orcid=0000-0002-2864-2377]{Margaret Garcia}
\affiliation{School of Sustainable Engineering \& the Built,  Environment, Arizona State University, Tempe, AZ 85281, USA}
\email{M.Garcia@asu.edu}
\author{Aaron Deslatte}
\affiliation{Paul H. O'Neill School of Public and Environmental Affairs, Indiana University, Bloomington, Indiana, USA, USA}
\email{adeslatt@iu.edu}
\author{Elizabeth A. Koebele}
\affiliation{Department of Political Science, University of Nevada, Reno, Nevada, USA}
\email{ekoebele@unr.edu}
\author{George M. Hornberger}
\affiliation{Department of Civil and Environmental Engineering, Vanderbilt University, Nashville, TN, USA}
\email{george.m.hornberger@vanderbilt.edu}
\author[orcid=0009-0001-1263-1251]{Adam Wiechman}
\affiliation{Princeton University, Princeton, NJ, 08544, USA}
\email{aw9050@princeton.edu}
\author[orcid=0000-0002-0138-8655]{John M. Anderies}
\affiliation{School of Human Evolution and Social Change and School of Sustainability, Arizona State University, Tempe, AZ 85281, USA}
\email{m.anderies@asu.edu}
\author{Jesse Barnes}
\affiliation{Department of Political Science, University of Nevada, Reno, Nevada, USA}
\email{jesse.barnes2@utsa.edu}
\author{Sara Alonso Vicario}
\affiliation{School of Sustainable Engineering \& the Built,  Environment, Arizona State University, Tempe, AZ 85281, USA}
\email{salonsov@asu.edu}
\author{Koorosh Azizi}
\affiliation{School of Sustainable Engineering \& the Built,  Environment, Arizona State University, Tempe, AZ 85281, USA}
\email{kazizi1@asu.edu}

\keywords{water supply $|$ institutional fit $|$ infrastructure $|$ robustness $|$ dynamic fitness}

\begin{abstract}
In an era of accelerating change, urban water infrastructure systems increasingly operate outside of their design conditions, putting new pressure on systems’ institutional designs to weather emerging challenges. Water management institutions must therefore be designed to exhibit “dynamic fitness,” defined by anticipatory capacity and responsiveness. However, we do not yet understand the specific features of institutional design that enable dynamic fitness, especially in relation to the diverse biophysical characteristics of systems that such fitness is contingent upon. We advance research on dynamic fitness in the context of urban water supply systems by drawing on 35-year data sets of stressors and responses for 16 U.S. urban water utilities using archetype analysis. Here we find that institutional archetypes capable of coping with higher biophysical complexity invest in both information processing capacity and response diversity. While dynamic fitness comes at a cost, balance between information processing capacity and response diversity promotes efficiency, which can be expanded through polycentric regional institutional structures that facilitate information sharing. Lastly, careful consideration should be given to tradeoffs across levels of governance, as institutional structures that facilitate dynamic fitness at the utility level may reduce the control and flexibility of higher levels of governance.
\end{abstract}

\section{Introduction}

Infrastructure – whether related to natural resource provision, transportation, telecommunications, or a myriad of other sectors –  is foundational to modern human well being, agency, and productivity. Even a few hours without functioning infrastructure can become an emergency and days a crisis. The robustness of infrastructure, or its ability to preserve performance in the face of specific disturbances \citep{Carlson2002, Rodriguez2011}, is therefore critical, especially in the modern era. Many infrastructure systems were largely designed for a world of environmental stationarity while the Anthropocene is characterized by accelerating change \citep{Chester2021, Steffen2015}, often producing more dynamic and extreme stressors that challenge infrastructure robustness.

There are three fundamental approaches to enhancing infrastructure system robustness: modifying the resource system structure, networking flows, and feedback control \citep{Anderies2015}. These approaches are integrated into the design of infrastructure systems as storage capacity, transport and conveyance structures, and control systems that detect change and trigger action, respectively. Crucially, these approaches pertain to more than just the physical components of infrastructure systems (e.g., dams, aqueducts, and sensors). The operational processes that facilitate flexible use of physical assets over short time scales, as well as the political and economic processes that govern investment in both maintenance and change over longer time scales, also contribute to enhancing robustness by enabling infrastructure systems to respond to a broad range of environmental and socioeconomic conditions \citep{Wiechman2024}. Consequently, robustness in the current era requires that the institutions structuring political and economic feedback pathways enable adaptation and transformation quickly enough to keep pace with change.

The critical role of the social and political processes underlying infrastructure highlights the importance of institutional design, or the design of arrangements of rules and norms that guide behavior and interactions among actors in a system, assign tasks and responsibilities, and allocate risks, benefits, and costs \citep{Koppenjan2005}. As much as we might hope, there are no panaceas in institutional design \citep{Ostrom2007}, and even the applicability of broad design principles is contingent on characteristics of the system and its environment \citep{Baggio2016,Ostrom1990}. The concept of “institutional fit” describes the alignment between institutional arrangements and the system to which they apply \citep{Ostrom1995, Young2002}. There are many aspects of institutional fit, including functional, temporal, and spatial fit \citep{Cox2012,Ekstrom2009}. In general, institutions are assumed to fit their context if they enable information processing and response actions that allow a system to persist in the face of the various types and magnitudes of disturbances it experiences (i.e., if they are robust). Recent scholarship moves from this static definition of fit to a concept of dynamic fitness, wherein responsiveness and anticipatory capacity enable robustness in a changing environment \citep{Moore2024}. In other words, institutional fit is a measure of the capacity of managing a stable variation regime, while dynamic fitness enables tracking changes in the environment. The concept of dynamic fitness prompts two broad questions for the design of infrastructure systems: 1) What features of institutional design enable dynamic fitness? 2) On what characteristics of an infrastructure system and its environment (biophysical) are these design features contingent?

We investigate these broad questions empirically in the context of urban water supply systems as they epitomize the pressures of environmental and social change on infrastructure systems \citep{Krueger2022, Rockström2014}. Approaches to designing urban water supply infrastructure for robustness include physical components that are quickly deployable such as storage, diverse supply portfolios, redundancy, sensors, and control systems \citep{Garcia2020, Gilrein2019}. At a longer time scale, political and economic feedback control processes may include adoption of new governance and management practices, as well as adapting physical infrastructure and its operation \citep{Garcia2019, Hornberger2015, Wiechman2024}. The specific details and timing of these actions in any given system are understood to be influenced by a combination of factors \citep{Azizi2024, Baggio2016}, such as the degree and types of socio-environmental stressors on the water supply system, its physical and institutional structure \citep{Garcia2019, Hornberger2015, Treuer2017}, and the way in which this structure produces, processes, and acts on information about stressors \citep{DeslatteAdams2024, Wiechman2024}. Consequently, water management authorities exhibit a spectrum of reactive to proactive and potentially transformative actions in response to short-term variability and longer-term systems change \citep{Rudnick2025}. Here, we focus specifically on understanding how a given system’s infrastructure and environment, which we collectively refer to as biophysical characteristics, and its governing institutions interact to facilitate dynamic fitness \citep{Garcia2019, Hornberger2015, Wiechman2024}.

In order to characterize and assess a system’s biophysical and institutional components and their interactions, we leverage the concept of archetypes, which depict essential features of a system that contribute to a phenomenon \citep{Magliocca2018}, based on analysis of recurrent patterns across multiple cases \citep{Oberlack2019}. As a method of analysis, archetypes offer a balance between generality and specificity, enabling a general description of the problem without losing important contextual features \citep{Eisenack2006}. A range of quantitative (e.g. cluster analysis, machine learning algorithms, meta-analysis) and qualitative (e.g. qualitative classification, qualitative comparative analysis, expert assessment) methods have been applied to identify and analyze archetypes \citep{Aggarwal2023, Sietz2019}. Here, we abductively develop biophysical and institutional archetypes of urban water systems to identify high level characteristics that influence their dynamic fitness. This use of archetypes facilitates synthesis across case studies to build general knowledge at an intermediate level of abstraction. Guided by this approach, we address three specific research questions with the goal of learning how to better design infrastructure systems for dynamic fitness: 

\begin{enumerate}
    \item What archetypal configurations of biophysical and institutional systems are likely to contribute to dynamic fitness in urban water supply cases? 
    \item What patterns emerge in the combinations of these archetypes and the observed dynamic fitness?
    \item What insight do these patterns give us in navigating infrastructure and institutional design in the face of environmental change?
\end{enumerate}

To answer these questions, we synthesize data collected from 1990 to 2024 on sixteen urban water utilities in the U.S. to characterize  their biophysical and institutional systems and to assess fitness \citep{Deslatte2022}. We then use abductive qualitative classification to identify and evaluate archetypes to explain why some urban water supply systems respond to stress and crises with adaptive changes while others experience repeated periods of poor performance.   

\section{Conceptual Foundations}

\begin{figure}[ht]
    \centering \includegraphics[width=1\linewidth]{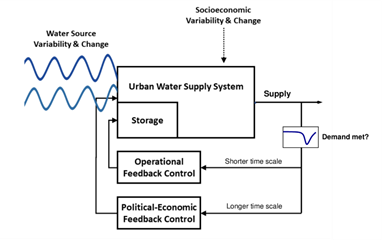}
    \caption{Conceptual figure illustrating how urban water supply systems manage variability in both water supply sources and socioeconomic conditions through passive robustness strategies such as drawing on multiple water sources and storage and through active response via feedback control. Adapted from \cite{Anderies2015}.}
    \label{fig:Fig1}
\end{figure}

The aggregate effect of infrastructure and institutional design on the performance of an urban water system or any infrastructure system under environmental variability and change can be framed as a robust control problem. The concept of robust control from modern control theory illustrates how urban water and other infrastructure systems process and act on information to respond to environmental changes \citep{Levin2022, Ostrom2011}. Water supply infrastructure robustness is enhanced when it is designed to manage supply and demand variability through networking multiple flows from diverse supply portfolios and storage that saves water from periods of high availability for use in periods of low availability \citep{McDonald2014, Padowski2012}. Robustness is also enhanced through active feedback control (Fig. \ref{fig:Fig1}). Over short time periods, water supply managers switch between water sources and shift reservoirs and treatment plants between operational modes by following established protocols (operational loop, Fig. \ref{fig:Fig1}). Over longer time periods, actors coordinate in political and economic processes to change the infrastructure, its operating rules, or the institutions governing management when observed or projected system performance is insufficient (political-economic loop, Fig. \ref{fig:Fig1}). This set of processes is far from an ideal feedback controller as it involves contested goals, imperfect knowledge, and coordination within and across multiple management organizations \citep{Anderies2019}. 


Feedback control is a powerful tool to maintain performance under variable and uncertain conditions \citep{Anderies2013}. However, enhancing a system’s robustness to a known set or frequency of disturbances may also induce fragility – defined as sensitivity of system performance to inputs – to other sets of disturbances, undermining broader system function \citep{Csete2002}. These tradeoffs are both inevitable and malleable, particularly in the face of uncertain future conditions. The design of policies and institutions shapes the conditions under which the system is robust, and higher-level institutions that enable a system to dynamically shift between policies increases the range of conditions in which a system is stable \citep{Anderies2019}. The dynamic fitness of institutional design is the extent to which it enables the actors in the system to mobilize knowledge to anticipate potential classes of changes, as well as the capacity to respond effectively if the need for a shift is detected. Critically, anticipation is not prediction (e.g. minimization of forecast error); instead anticipation orients attention to general classes of events \citep{Boisot2011}. This underscores the importance of institutional design aspects from the level of operational rules that specify day-to-day actions to higher-level collective institutional architecture that shapes how rules can be changed and enables and constrains local-level information processing and action \citep{DeslatteAdams2024}.

\subsection{The Biophysical System: Natural and Built Infrastructure}

Urban water systems mobilize the natural infrastructure of watersheds (surface water) and aquifers (groundwater). Surface water sources are characterized by variation across daily, seasonal, and decadal time scales \citep{Dettinger2000, Singh2021, Yao2020}. These frequency signatures vary regionally \citep{Ho2017}, with variation driven by hydrometeorological dynamics \citep{Peixóto1984, Yao2020}. How hydrometeorological variability propagates to streamflow variability depends on the temporal interaction between supply (precipitation) and demand (potential evapotranspiration), as well as the extent of soil and groundwater storage which enables the watershed to retain and release water over time \citep{Alonso2024, Massmann2020, Rice2017}. Groundwater varies across seasonal to decadal time scales, with the variation driven by hydrometeorology and shaped by hydrogeological characteristics \citep{Güntner2007, Li2015}. Further, some surface and groundwater sources exhibit trends attributed to land use change, water withdrawals, and climate change \citep{Humphrey2016,  McCabe2016, Rice2016, Saki2023}. Observed and projected changes in precipitation and temperature caused by climate change are altering these patterns across time scales in the Anthropocene \citep{Marvel2023, Swain2025}. 

As described above, urban water supply systems smooth temporal variability in supply sources through storage and by mixing a portfolio of water supply sources \citep{McDonald2014, Padowski2012}. For example, reservoirs along with their operating rules smooth high-frequency variability in water supply (e.g. daily, seasonal), effectively reducing the variability that the rest of the system must manage \citep{Garcia2020}. Critically, the political and economic system does not need to manage the full variability of local hydrology but instead benefits from prior investments in building and maintaining infrastructure. As climate change drives shifts in hydrology, the burden on the political and economic system may increase given limitations of the response capacity of infrastructure and its operations. The residual variability not controlled by the infrastructure system contributes to the biophysical complexity.

In addition to using infrastructure to manage temporal variability, urban water supply systems include water conveyance infrastructure, water and wastewater treatment plants, and water distribution and wastewater collection networks. Such infrastructure  is often characterized as having high inertia \citep{Davis2010, Williams2021}, or a “highly constrained ability to adapt to changing internal and external conditions'' due to both the challenge of rewiring the system from within and the large scale of such systems \citep{Chester2021}. Characteristics that contribute to high inertia include centralized components that serve the full demand, centralized distribution and collection network structures, long design life, and interdependence with other components or systems \citep{Helmrich2023}. While more flexible and lower inertia designs are feasible (e.g., decentralized distribution, modular storage and treatment \citep{Turlington2017}), U.S. urban water supply systems are largely centralized and have high inertia. Overcoming this inertia to change the structure or function of infrastructure systems requires substantial resources and time, so the degree of inertia shapes the effective solution space over short time scales. Urban water systems with more diverse water supplies or that otherwise have a higher coordination burden have higher biophysical complexity. In other words, the biophysical complexity is the combination of the variability that passes through the infrastructure system and the complexity of management the infrastructure requires. 

\subsection{Institutional Design}

In the context of urban water management, institutional designs consist of rules, norms, and shared strategies that support infrastructure operation, maintenance, and investment \citep{Daems2024, Ostrom2009}. At the regional level, multiple urban water systems may interact and influence one another, and state and regional entities may enable or constrain their actions. Urban water systems can therefore be described as operating in polycentric governance arrangements featuring multiple centers of decision-making authority working under a common set of rules \citep{Aligica2012}.  Such arrangements are thought to emerge to effectively govern complex social or environmental systems because they leverage the benefits of both context-specific, decentralized governance to support competition and bottom-up information flow, as well as of centralized, or at least coordinated, governance under common rules that make governing simpler and more predictable \citep{Daems2024}. The degree of (de)centralization of a polycentric arrangement can influence information flows and enable or constrain response diversity. 

  \begin{figure*}[ht]
    \centering \includegraphics[width=0.9\linewidth]{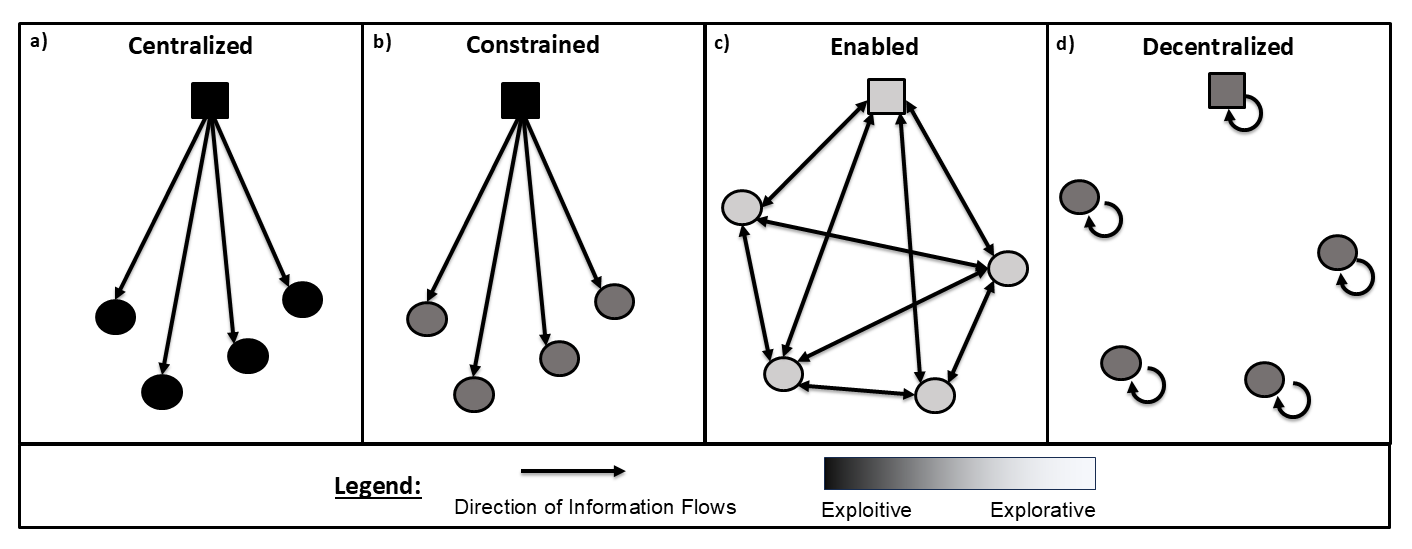}
    \caption{Institutional configurations where each circle is an urban water system and the square is a regional or state level entity: a) centralized, where information flows in one direction and decisions are largely top-down; b) constrained, where information flows in one direction and the decisions on what outcomes to achieve are top-down but lower level make implementation decisions; c) enabled, where authority for decision making is shared across the network, information flows are bidirectional both horizontally and vertically; d) decentralized, where authority for decision making is shared across the network but connectivity and information flows are limited. }
    \label{fig:Fig2}
\end{figure*}

Equally important are the utility-level institutional design choices to impose institutional (decision) costs and create flexibility. They shape urban water systems’ ability to process information and translate information into response actions \citep{Daems2024}. Institutional costs arise from the time, effort and expense of complying with regulatory, contractual or information-processing requirements \citep{Cox2012, DeCaro2017, Ostrom1995} and affect the proportional response to new error signals. For instance, institutional designs may facilitate specific types of information collection or use, which can affect anticipatory capacity by increasing information costs, creating veto points, or requiring prerequisites for action \citep{Mesdaghi2022}. Institutional designs can also expand or restrict the participants and the scope of goals by influencing the diversity of expertise or experience involved in decision-making \citep{DeslatteKoebele2024}. Higher institutional costs can serve to reduce the risk of surprises through stable, established policy and management processes \citep{Deslatte2023} but also risk inhibiting information flows necessary for experimentation and innovation \citep{DeslatteKoebele2024}.

The flexibility of the institutional design of an infrastructure system influences administrative discretion and the reflexivity of decision-making. Institutional designs with more flexibility can facilitate more exploratory (i.e., epistemic) evidence-gathering \citep{Veissière2020}, while less flexible institutions may prescribe more exploitative (i.e., pragmatic) evidentiary effort. While more flexible institutional designs are generally viewed as advantageous, flexibility can also be inefficient (e.g., analysis paralysis), reduce predictability, and even lead to disaster if safeguards are absent. Thus, through institutional costs that condition the flexibility of decision-making, institutional designs may function to enable or constrain feedback pathways and their underlying attention dynamics.

Considering both the design of the polycentric governance arrangement in which an urban water system is nested, as well as the utility-level institutional design, can provide insight into the overall institutional configuration in which a system operates. For example, fully centralized institutional configurations display top-down decision making and allow minimal flexibility (Fig \ref{fig:Fig2}a). They are effective in low complexity environments and have low institutional costs. However, their structure limits information flows and emphasizes exploiting the information over exploring, limiting error-correction capacity when the operating environment changes. Highly constrained institutional configurations establish preconditions for action, narrow or restrict the goals and limit evaluative criteria but allow some flexibility in implementation \citep{DeCaro2017} (Fig \ref{fig:Fig2}b). They tend to display more top-down decision making and less ground-up information feedback. As complexity increases, constrained designs restrict the diversity and spontaneity of decision making \citep{Daems2024}. These institutional designs may limit flexibility by restricting the evaluative criteria (i.e. cost-efficient versus equitable) or options for responding to new problems, increasing the inertia of the institutional system. Highly enabled configurations also establish preconditions for action but may allow for more varied goals and evaluative criteria, facilitating greater diversity in policy or programmatic responses to problems and allow bidirectional information flows across the network (Fig \ref{fig:Fig2}c). Thus, enabled designs impose higher institutional costs but also open the problem and solution spaces (e.g., high flexibility) for exploring and exploiting the dynamics of evolving resource system components \citep{DeslatteKoebele2024}. Fully decentralized institutional configurations delegate authority, enabling information collection and flexibility, and resulting in high potential response diversity (Fig. \ref{fig:Fig2}d). However, fragmentation limits information flows between urban water systems leading to redundant efforts and inference constraints. Thus, institutional configurations impose tradeoffs between ability to manage complexity, their efficiency, and the degree of direct control they offer at different levels. 

  \begin{figure*}
    \centering \includegraphics[width=0.9\linewidth]{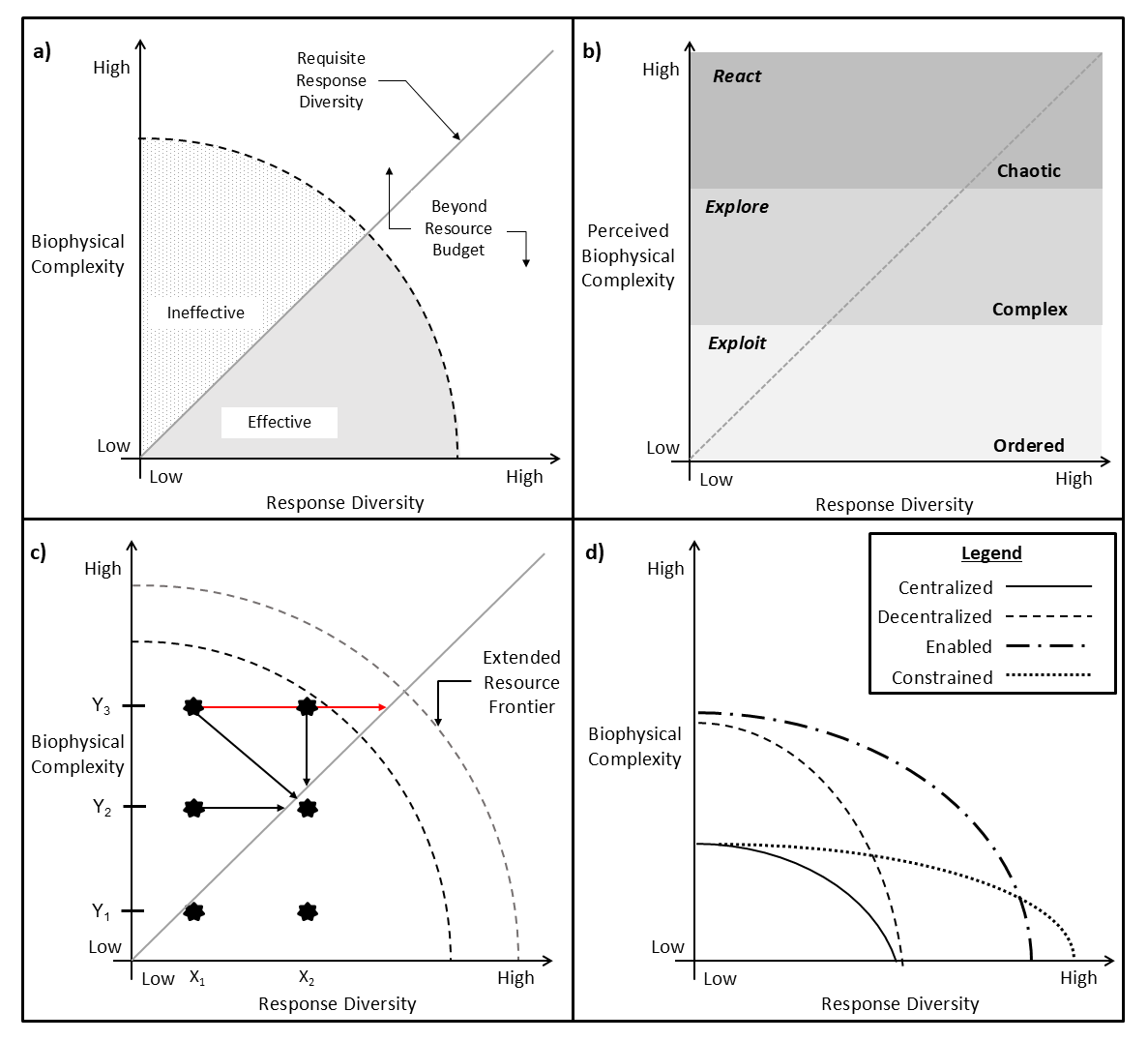}
    \caption{a) An adaptation of the Ashby Space diagram illustrating the requisite response diversity as a function of biophysical complexity where responses are also constrained by available resources (adapted from \citep{Boisot2011}); b) Three regimes of biophysical complexity (labeled in bold) are projected onto the Ashby space with the types of knowledge mobilization effective in each regime indicated in italics; c) This diagram plots the potential responses of two entities with different levels of response diversity (X) to varying degrees of biophysical complexity (Y). d) This diagram maps institutional configurations to the Ashby Space where each configuration is effective inside of the illustrated boundaries.}
    \label{fig:Fig3}
\end{figure*}

\subsection{The Interaction between Biophysical Complexity and Institutional Design}

Given our definition of dynamic fitness, a system has fitness if the institutional design enables the mobilization of knowledge to anticipate challenges and invests in the capacity to respond in a timely and effective manner to the range of encountered and envisioned problems. To better explain how anticipatory capacity and response capacity work in tandem, we draw on Ashby’s Law of Requisite Variety \citep{Ashby1968} and recent developments building upon this law \citep{Boisot2011}. Ashby’s Law of Requisite Variety states that only variety in response can reduce the variety in outcomes relative to the variety in the biophysical environment (Fig \ref{fig:Fig3}a). However, not all variety in the environment is equally important. With respect to a given system and set of goals, some variety is effectively noise. An organization’s ability to isolate the signal and anticipate challenges and consequences of response actions, is a function of its ability to mobilize knowledge to understand system dynamics. Where biophysical dynamics can be understood as ordered, decisions are routine and exploitive information processing suffices; where they are understood as complex, decisions are strategic and explorative information processing is required; and where they are understood as chaotic, information processing is ineffective and decisions are reactive (Fig \ref{fig:Fig3}b). 

At a given moment, an entity has a specific degree of response diversity available, though further response diversity may be incipient \citep{Moore2024}. For example, take two institutional designs: one with response diversity $X_1$ and the other with response diversity $X_2$ (Fig \ref{fig:Fig3}c). At level $Y_1$ of biophysical complexity, both are effective institutional designs that have more than the requisite diversity. As biophysical complexity increases to $Y_2$, institutional design $X_2$ remains effective while the entity with design $X_1$ must act to remain effective. As $X_1$ is far from its resource boundary, it can add response diversity via trial and error (moving right and mobilizing incipient response capacity) before exhausting its resources. However, at level $Y_3$ of biophysical complexity, neither design is effective and the proximity of the resource boundary means that trial and error will not suffice unless the resource boundary can be extended. If the entities do not have the capacity to engage in exploratory information processing to accurately infer the new biophysical dynamics, they may make potentially maladaptive simplifying assumptions (moving down). Alternatively, if they have the capacity, they may apply explorative information processing to improve inference and adopt additional targeted response diversity (moving diagonal). This mapping also illuminates four potential failure modes of institutional design under increasing biophysical complexity: 1) iteratively responding through trial-and-error without an effective guiding mental model and exhausting resources (moving horizontally to the right); 2) ignoring complexity and responding with existing routines (moving down), 3) trying to force complexity into existing mental models, risks mischaracterization and mal-adaptive response (moving diagonally down and to the right with inaccurate inference); and 4) optimizing for efficiency (minimizing resource boundary) and limiting incipient capacity (also referred to as the Icarus paradox \citep{Miller1992}). Each institutional configuration consists of a distinct combination of investments in anticipatory capacity and response diversity that shapes the range of biophysical complexity where they can be effective (Fig \ref{fig:Fig3}d). 

Despite theoretical development at the intersection of biophysical complexity and institutional design, there is limited empirical research to date on how this interaction manifests and affects fitness in the context of large scale infrastructure systems. This gap motivates our work on archetypal configurations of the biophysical and institutional system and ability of urban water supply systems to keep pace with changing conditions.

\section{Materials and Methods}

While the aim of designing systems for dynamic fitness is forward looking, we cannot empirically evaluate future performance, we can only learn from the past. Therefore, we look back to inform the path forward by synthesizing data from 16 U.S. large urban water suppliers over the period from 1990 to 2024.

\subsection{Cases and Data}

\begin{figure*}
    \centering \includegraphics[width=0.9\linewidth]{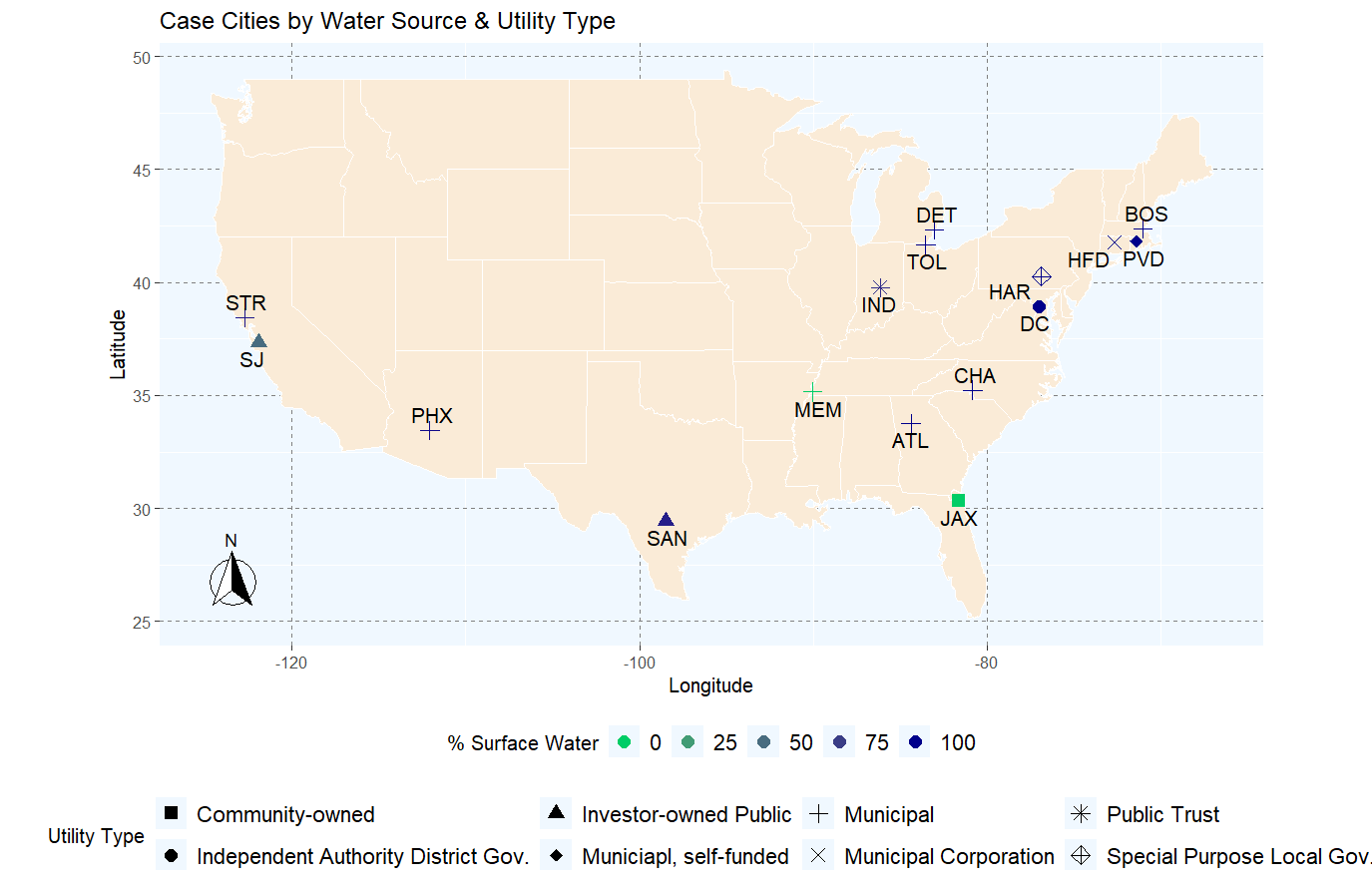}
    \caption{Case study urban water suppliers with the percent surface water in their supply portfolio indicated by color and the utility type indicated by shape.}
    \label{fig:Fig4}
\end{figure*}

We selected 16 U.S. large urban water supply utilities that are representative of the diversity of water management structures, climatic conditions, community characteristics, and water supply portfolios \citep{Azizi2024} (Fig. \ref{fig:Fig4}). These cases were selected from an initial pool of 197 large U.S. cities  following the method documented in \citep{Deslatte2022}. The diversity in this case set enables the development of archetypes applicable to the broader class of large U.S. urban water utilities. The characteristics of the 16 utilities are summarized in Figure 3 below, and include the following: Atlanta, GA (ATL); Boston, MA (BOS); Charlotte, NC (CHA); Detroit, MI (DET); Harrisburg, PA (HAR); Hartford, CT (HFD); Indianapolis, IN (IND); Jacksonville, FL (JAX); Memphis, TN (MEM); Phoenix, AZ (PHX); Providence, RI (PVD); San Antonio, TX (SAN); San Jose, CA (SJC); Santa Rosa, CA (STR); Toledo, OH (TOL); and Washington, DC (DC). Further description of the cases can be found in the supplementary information (S1). For each city, we aggregated extensive data on the hydrology and climate, infrastructure design, institutional design and a case history covering, at a minimum, the period from 1990 to 2024. 

\subsubsection{Hydroclimatic and Infrastructure Metrics}

To characterize the time scales and magnitude of water source variability and the infrastructure design, we use three metrics: the Köppen aridity index, seasonality from the Köppen-Geiger climate types, and the Gini Index of Water Supply Diversity. The Köppen aridity index is the ratio of the mean annual precipitation and mean annual temperature plus a constant \citep{Quan2013}. The Köppen-Geiger climate types place locations on a three tiered classification scheme using long-term monthly precipitation and temperature \citep{Peel2007}. Part of this classification addresses seasonality specifically if a location experiences a dry or a cold (below 0 ℃) season. Here we classify all cities as having seasonality if they experience either a dry or a cold season. Lastly, following \citep{Gonzales2017}, we compute the Gini Index of Water Supply Diversity as a function of the average annual volume of water provided by each of a city’s water sources. This metric captures the extent to which a utility is using multiple water sources to manage variability. The Gini Index of Water Supply Diversity indirectly measures the complexity of managing the infrastructure, as more distributed and decentralized systems with a greater number of sources will have a higher index value.

To use these metrics to inform archetypes, we first identified thresholds for aridity and source diversity to divide the metric range into classes. For aridity all values above 15 were classified as humid, with those below classified as dry (encompassing arid and semi-arid). For source diversity, we identified two categories of infrastructure complexity and set a fuzzy threshold, with gini index below 0.3 with full membership in "Simple" and above 0.5 with full membership in "Complex." As seasonality was already a binary variable no additional processing was needed. The fuzzy threshold allows for qualitative data to inform edge cases. 

\subsubsection{Institutional Data}

Given the multilevel nature of governance arrangements, data to develop institutional archetypes was assembled in three steps. First, we used prior work to code the “Institutional Grammar” (IG) or syntactic components of state laws, local ordinances and regulatory rules, to identify boundary and choice conditions related to utility rate-setting and investment \citep{Crawford1995, Deslatte2022}. This allows us to begin assessing the institutional costs and flexibility of a design via the preconditions for taking action (empirically labeled “institutional dependencies,” following \citep{Deslatte2023}), the diversity of actors involved, and any formally specified response options \citep{DeslatteKoebele2024}. Second, we used the IG and 2012, 2017, and 2022 State Scorecard data from the Alliance for Water Efficiency, a nonprofit organization which tracks state water policies in North America, to identify whether states require drought and conservation planning. While planning processes do not guarantee reflexivity or the use of more diverse knowledge types, some empirical evidence suggests they do encourage the use of a broader timescale for managing risks and a wider array of scientific, technical or experiential knowledge will be developed and mobilized \citep{DeslatteAdams2024, Deslatte2025}. We use this to assess the enabling characteristics of the institutional design (i.e., whether drought contingency plans are required and updated regularly) \citep{DeslatteAdams2024}. The third step involves determining the constraining characteristics of a design, based on whether state regulatory oversight of utility rate-setting is required. Here again, we draw from empirical support that centralized regulatory oversight of rate-setting, which typically takes the form of a quasi-judicial evidentiary process, tends to impose formal restrictions on the types of information, participation, and evaluative criteria used to review utilities’ infrastructure needs \citep{Deslatte2023}. Thus, constrained designs impose high institutional costs but provide less flexibility to utility managers compared with designs that impose planning requirements but largely leave infrastructure investment decisions to local control \citep{Daems2024, Deslatte2025}. 

\subsubsection{Response Data}

Given the need to evaluate dynamic fitness empirically based on past observations, we focus our analysis of response on observed crisis level disruptions that either lead to performance impacts or where response was taken in anticipation of projected impacts. To characterize responses systematically, we created timelines of infrastructure and policy events related to urban water supply for each city from the development of the supply system to 2024 by drawing on utility plans, reports and websites, municipal, state and federal agency reports, plans and legislation, environmental NGO documents, local historical organizations reports, and local news articles \citep{Garcia2025}. Here we used the period from 1990 to present as documentation availability is more variable prior to 1990. A broad range of crises were included to capture the full range of operational challenges facing water suppliers but soley drought crises are presented here (Table \ref{tab:Tab1}). Each crisis event response was coded in terms of timeliness, category of response, governance level(s) and effectiveness. Full details are provided in S2. 

To provide a check on the coded response data and to enable further analysis, we prepared case narratives that detail the infrastructure and water management context, the characteristics, impacts of and response to crisis events, and other infrastructure or management changes over the study period that may impact the utilities vulnerability to hydroclimatic extremes and its ability to respond to crisis events.

\begin{table*}[!t]
  \centering
     \caption{Summary of drought crises events and response characteristics. Infrastructure abreviated as inf.}
    \begin{tabular}{cccccccc}
    \toprule
       \textbf{City}  & \textbf{Year} & \textbf{Crisis Type} & \textbf{Response Type} & \textbf{Timing} &  \textbf{Level} & \textbf{Effective} \\
       \midrule
       Atlanta & 2007 & Drought & Planning, conservation & Prompt & Utility, State & Yes \\
       Boston & 1988 & Drought & Planning, conservation, inf. & Prompt & Utility, State & Yes \\
       Charlotte & 2002 & Drought & Inf., planning, authority & Prompt & Utility, State & No \\
       Charlotte & 2008 & Drought & Planning, conservation & Prompt & Utility, Local, State & No \\
       Charlotte & 2012 & Drought & Planning, authority & Prompt & Utility, Local, State & Yes \\
       Harrisburg & 1991 & Drought & Inf. & Medium & Utility & No \\    
       Harrisburg & 1999 & Drought & Inf., policy & Prompt & Utility, State & No \\   
       Harrisburg & 2002 & Drought & Policy, planning & Prompt & Utility, Local, State  & Unknown \\   Indianapolis & 1988 & Drought & Planning & Delayed & State  & Yes \\      
       Indianapolis & 1988 & Drought & Planning, inf. & Prompt & Utility, Local, State  & Yes \\      
       Phoenix & 2003 & Drought & Planning, conservation, inf. & Prompt & Utility, Local, State  & Yes \\     
       San Antonio & 2011 & Drought & Planning, authority, inf., cons. & Prompt & Utility, Local, State  & Yes \\    
       San Jose & 2007 & Drought & Policy, conservation & Prompt & State  & No \\    
       San Jose & 2014 & Drought & Policy, conservation & Prompt & Utility, Local, State  & Yes \\  
       Santa Rosa & 2014 & Drought & Policy, conservation & Prompt & Utility, Local, State  & Yes \\
    \bottomrule 
    \end{tabular}
\label{tab:Tab1}
\end{table*}

\subsection{Archetype Development and Analysis}

We abductively develop two sets of building block archetypes. Building block archetypes focus on sub-case level phenomena and describe a component or process that may be observed across multiple cases \citep{Eisenack2021, Oberlack2019}. Each case may contain multiple building block archetypes. From our conceptualization of the system as a case of robust control (Fig. \ref{fig:Fig1}), we identify two building blocks: the biophysical system inclusive of the hard infrastructure and operational feedback loop and the institutional system inclusive of the political-economic feedback loop. The data sets described in sections 3.1.1 and 3.1.2 were selected to inform these archetypes based on the theory described in section 2. We then review the attributes of the biophysical and institutional systems to develop the archetypes. By decomposing cases into these building blocks, we can identify the alternate structures of each block, pattern the combination of blocks, and assess the link between those combinations and responses to disruption events. 

\section{Results}

\subsection{Archetypes}

The analysis of aridity, seasonality and water supply diversity across the sixteen cases resulted in the identification of four biophysical archetypes, encompassing hydrology and infrastructure, seen in Table \ref{tab:Tab2} 1) easy hydrology (humid and no seasonality) and simple infrastructure; 2) seasonal hydrology and simple infrastructure; 3) seasonal hydrology and complex infrastructure; and 4) dry and complex infrastructure. Some possible combinations such as humid, no seasonality and complex infrastructure were not observed. This is expected as such combinations would not support the requisite institutional diversity. Cities with dry climates and complex infrastructure with and without seasonality were combined to a single Dry and Complex Infrastructure archetype, as water scarcity is the primary driver of water management approaches in those cities.

\begin{table*}
  \centering
     \caption{Biophysical Archetypes and Supporting Data.}
    \begin{tabular}{cccccl}
    \toprule
       \thead{City}  & \thead{Köppen\\ Aridity} & \thead{Köppen-Geiger\\ Climate Type} & \thead{Gini\\Index} & \thead{Infrastructure Class} &  \thead{Biophysical Archetype} \\
       \midrule
       Atlanta & 24.8 & Cfa & 0.45 & Simple & 1) Easy hydro., simple inf.\\
       Boston  & 28.2 & Dfb & 0.48 & Complex &  3) Seasonal hydro., complex  inf.\\
       Charlotte  & 22.6 & Cfa & 0.32 & Simple  &  1) Easy hydro., simple  inf.\\
       Detroit  & 18.4 & Dfb & 0.10 & Simple  &  2) Seasonal hydro., simple  inf.\\
       Harrisburg  & 24.3 & Dfa &  0.14 & Simple  &  2) Seasonal hydro.,  simple inf.\\
       Hartford  & 29 & Dfb &  0.36 & Complex &  3) Seasonal hydro., complex inf.\\
       Indianapolis  & 24.4 & Dfa & 0.79 & Complex &  3) Seasonal hydro., complex inf.\\ 
       Jacksonville  & 24.3 & Cfa & 0.00 & Simple  &  1) Easy hydro., simple inf.\\ 
       Memphis  & 25.8 & Cfa &  0.00 & Simple  &  1) Easy hydro., simple inf.\\ 
       Phoenix  & 3.6 & BWh & 0.66 & Complex &  4) Dry, complex inf.\\
       Providence  & 26.7 & Dfa & 0.00 & Simple  &  2) Seasonal hydro.,  simple  inf.\\
       San Antonio  & 13.9 & Cfa &  0.58 & Complex &  4) Dry,  complex inf.\\
       San Jose & 10.6 & Csb &  0.58 & Complex &  4) Dry, complex inf.\\
       Santa Rosa  & 14.8 & Csb & 0.34 & Complex &  4) Dry, complex inf.\\
       Toledo  & 20.6 & Dfa &  0.00 & Simple  &  2) Seasonal hydro., simple inf.\\
       Washington & 22.6 & Cfa & 0.00 & Simple  &  1) Easy hydro., simple inf.\\
    \bottomrule 
    \end{tabular}
\label{tab:Tab2}
\end{table*}

Table \ref{tab:Tab3} details four abductively developed institutional archetypes based on their institutional configuration (centralized, constrained, enabled or decentralized), with respect to both their institutional costs and flexibility. Because we expect the degree of biophysical complexity to influence institutional design, Table 3 also hypothesizes the level of biophysical complexity that matches with each ‘ideal’ institutional archetype. Note that none of the cases fall into the Centralized archetype but that it can be observed globally, while Boston, Harrisburg, and Toledo, meet the decentralized definition because they are not subject to state regulatory oversight or mandated drought or conservation planning. Constrained designs such as Indianapolis, Providence, and San Jose, feature high institutional costs in the form of state regulatory oversight and low flexibility, characterized by limited reflexiveness or discretion in how problems are defined and addressed. Enabled designs like Phoenix, Detroit, and Charlotte, feature high institutional costs but also high flexibility, characterized by greater latitude to explore alternative explanations for environmental conditions and options for policy or management responses. 

\begin{table*}
  \centering
     \caption{Institutional Archetypes Characteristics and Sorting Criteria.}
    \small
    \begin{tabular}{clllll}
    \toprule
       \textbf{Attribute Type}  & \textbf{Attribute} & \textbf{1) Centralized} & \textbf{2) Constrained} & \textbf{3) Enabled} &  \textbf{4) Decentralized} \\
       \midrule
        Characteristics & Institutional Costs: & Minimum & Medium & High & Maximum \\
         & costs of coordinating, & & & & \\
         & negotiating, \& enforcing & & & & \\
         & agreements or policies & & & & \\
        Characteristics & Flexibility: greater & Minimum & Medium (goals, & High & Maximum \\
         & prerogative to explore & & actions \& evaluative & & \\
         & alternative options & & criteria are formalized) & & \\
        Expectations & Empirical Fingerprint & Few dependencies, & Many dependencies, & Many dependencies & Few dependencies, \\
         & (things we expect to & voids; consolidated &  few voids; highly & \& voids; moderately & many voids; min.\\
         & see in the archetype & districts; regional  & regulative; optional  & regulative; required & regulation or\\
          & category) & authorities. & planning. & planning. & planning. \\
        Expectations & Biophysical Complexity & Low & Low & High & High \\
    \bottomrule 
    \end{tabular}
\label{tab:Tab3}
\end{table*}

\subsection{Response Patterns }

Table \ref{tab:Tab4} summarizes the biophysical and institutional archetypes assigned to each city along with the city’s responsiveness to past drought crises. Per Table \ref{tab:Tab3}:, enabled and decentralized archetypes are hypothesized to be ideal for complex biophysical conditions but are not expected to lead to poor performance in simple biophysical conditions. Therefore, it is unsurprising to see cases, such as Atlanta, that exhibit these combinations have a history of prompt and effective drought response. Notably, we do not see evidence that the constrained institutional archetype, hypothesized to best fit simple biophysical conditions, is consistently paired with the simple biophysical archetypes 1 and 2. In contrast, we observe two instances (San Antonio and San Jose) where constrained institutions are found in cases with dry climates and complex infrastructure systems. The San Antonio, local government and water utility responded to the 2011 drought promptly and effectively and responses came from the utility, local government and state government. However, as San Antonio pumps groundwater from the Edwards Aquifer where years of dispute led to a comprehensive planning effort, it may effectively operate as an enabled system because it is required to participate in this planning process \citep{Votteler2023}. In contrast, San Jose, did not respond locally to the 2007-09 drought, though the state government did respond. (Notably San Jose is constrained while Santa Rosa is enabled because it has an investor owned utility and the regulatory regime in CA requires different oversight of private utilities.) During the 2012-16 CA drought, there was a 2014 city level water shortage declaration in San Jose which aligned with a state level resolution mandating urban conservation. All other response actions were at the state level, demonstrating the strong state role in drought response in San Jose as expected under a constrained institutional design.

\begin{table*}[!t]
  \centering
     \caption{Characteristics of crisis response by city along with biophysical and institutional archetypes. Bold indicates cities that experienced droughts crises.}
    \small
    \begin{tabular}{ccllll}
    \toprule
       \textbf{City}  & \textbf{Crisis} & \textbf{Response Summary} & \textbf{Response Level} & \textbf{Biophysical Archetype} &  \textbf{Inst. Archetype} \\
       \midrule
        Atlanta & Drought & Prompt, effective &  Utility, state & 1) Easy hydro., simple inf. & 3) Enabled \\
        Boston & Drought & Prompt, effective &  Utility, state & 3) Seasonal hydro., complex inf. & 4) Decentralized \\
        Charlotte & Repeated & Prompt ineffective, &  State first, later &  1) Easy hydro, simple inf. & 3) Enabled \\
         & drought & ultimately effective & utility, local & & \\
        Detroit & Financial &  Prompt, not effective  &  Utility, local, state &  2) Seasonal hydro., simple inf. & 3) Enabled \\
        Harrisburg & Repeated &  Initially delayed, &  Utility, local, &  2) Seasonal hydro., simple inf. & 4) Decentralized \\
         & drought & later prompt, effective & state & & \\
        Hartford & None  & N/A & N/A & 3) Seasonal hydro., complex inf.  & 3) Enabled \\
        Indianapolis & Drought &  Initially delayed, &  State first,  & 3) Seasonal hydro., complex inf. & 2) Constrained \\
         & & later prompt, effective & later utility, local & & \\
        Jacksonville & Hurricane &  Prompt, effective &  Local, state  &  1) Easy hydro., simple inf.   & 3) Enabled \\
        Memphis & None  &  N/A & N/A & 1) Easy hydro., simple inf. & 3) Enabled \\
        Phoenix & Drought  &  Prompt,  effective & Utility, local, state & 4) Dry, complex inf.   & 3) Enabled \\
        Providence & None  & Droughts experienced & N/A & 2) Seasonal hydro., simple inf. & 2) Constrained \\
         & & but no crises & & & \\
        San Antonio & Drought  &  Prompt, effective & Utility, local, state & 4) Dry, complex inf. & 2) Constrained \\
        San Jose & Repeated & Prompt w/ variable & Primarily state & 4) Dry, complex inf. & 2) Constrained \\
         & drought & effectiveness & & & \\
        Santa Rosa & Drought, & Prompt drought, &  Utility, local &  4) Dry, complex inf. & 3) Enabled \\
         & flooding & no flood response & state & & \\
        Toledo & Water quality  &  Prompt, effective & Utility, local & 2) Seasonal hydro., simple inf. & 4) Decentralized \\
        Washington & None & Droughts experienced & N/A  &  1) Easy hydro., simple inf. & N/A \\
         & & but no crises & & & \\
    \bottomrule 
    \end{tabular}
\label{tab:Tab4}
\end{table*}

\subsection{Learning from Crises}

Over the 1990-2024 study period, several water suppliers, including Charlotte, Harrisburg, Indianapolis and San Jose, experienced repeated drought crises. They illustrate dynamic fitness through changes in responsiveness and the effectiveness of response over time. These cases offer a window into how water suppliers learn from crises events, responses and impacts. We prepared narratives of three of these cases, Charlotte, Harrisburg, and Indianapolis, which represent three different biophysical and institutional archetypes (S3). Below we highlight key events and insights from these narratives. 

\begin{figure*}
    \centering \includegraphics[width=0.95\linewidth]{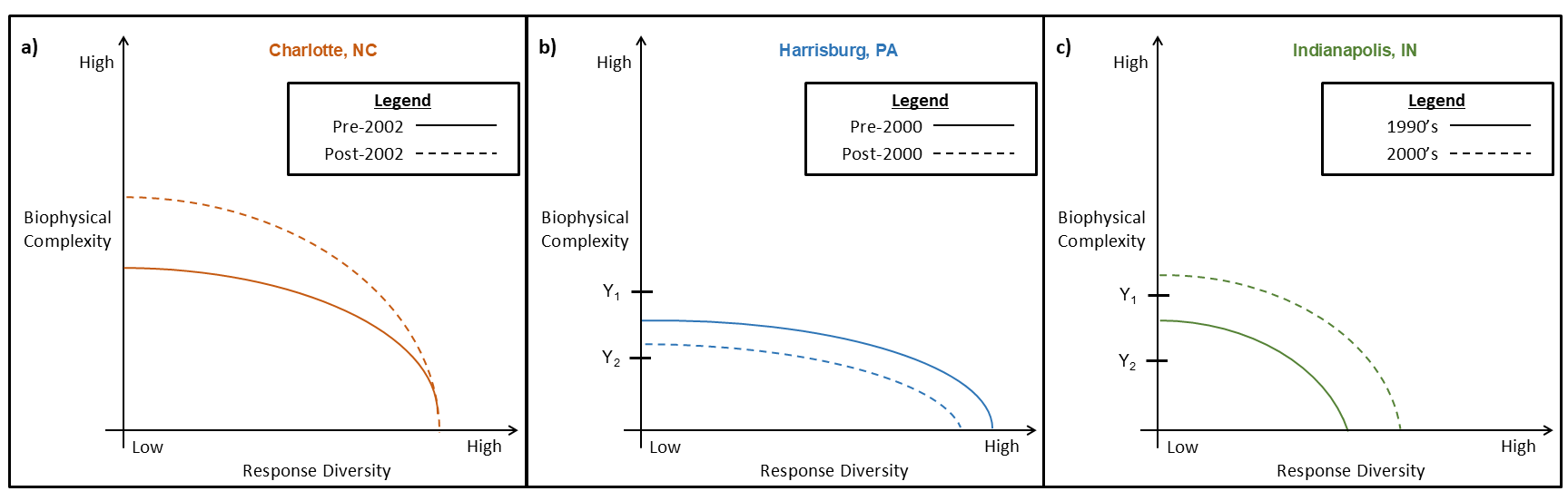}
    \caption{Evolution of institutional configurations over time plotted in the Ashby space showing the change in space where it each is effective: a) Charlotte, increased information processing capacity through state level investments in drought information collection and low flow protocols; b) in Harrisburg, investment in backup supply infrastructure shifted the biophysical complexity the institutions must respond to from $Y_1$ to $Y_2$, and over time a decreasing consumer base reduced resources available shrinking the effective space; c) in Indianapolis, investment in additional supply infrastructure shifted the biophysical complexity the institutions must respond to from $Y_1$ to $Y_2$ while institutional capacity increased by working through complex water quality challenges.}
    \label{fig:Fig5}
\end{figure*}

Over the study period, Charlotte faced three major droughts and experienced significant population and demand growth. The 1998-2002 drought led to record low reservoir levels across North Carolina, including the water supply reservoirs for Charlotte. Response to the severe drought was prompt at the state level and included the creation of new state level entities tasked with providing drought information and guidance on water resources management, and the development of an operational protocol for low streamflow. However, there were no utility led responses. Charlotte is characterized as an enabled institutional archetype and as having a simple biophysical environment (easy hydrology and simple infrastructure). Under these circumstances, a prompt local level response is anticipated if needed. Charlotte faced drought again in 2008 and 2012. Both state and local level responses were prompt and included conservation programs and water sustainability planning. It is possible that the state level investments in drought information and low streamflow protocols after the 2002 drought, in part, enabled prompt local and utility level responses in subsequent droughts (Fig. \ref{fig:Fig5}a). It is also important to note that the responses to the 2008 drought were not sufficient to make the system robust to the 2012 drought, but that later droughts (e.g., winter 2016-2017) did not rise to a crisis level. 

Harrisburg experienced three statewide drought emergencies during the study period: 1991-92, 1999, and 2002. In response to the 1991-92 drought, Capital Region Water, the Harrisburg water supplier, built the Susquehanna River Pump Station to serve as an emergency backup source if water levels at the DeHart Reservoir fall too low (Fig. \ref{fig:Fig5}b shift from $Y_1$ to $Y_2$). After the 1999 drought, the utility installed an automated water meter reading system to track demand more efficiently. After the 2002 drought, the state passed the Water Resources Planning Act, which created a statewide water resources plan and regional commissions to review local plans. The utility’s prompt response to 1991-92 and 1999 droughts is consistent with expectations for a decentralized institutional archetype and a simple biophysical environment (biophysical archetype 2, seasonal hydrology and simple infrastructure). Since 2002, no severe droughts have occurred, and the city and utility have been focused on financial and water quality challenges. While the metropolitan region of Harrisburg grew steadily during this period, the 1990s and early 2000s were a period of population decline in the city of Harrisburg where the majority of utility customers are located, exacerbating financial challenges (Fig \ref{fig:Fig5}b). This highlights the importance of availability of resources for local response. 

Indianapolis experienced two severe droughts from 1990-2024, a period which also saw multiple changes in water supply ownership and an expanding coverage area. A drought in the late 1980s led to only a delayed, state level response in the early 1990s. This is an anticipated effect of a constrained institutional archetype in a complex biophysical environment. However, we do not see repeated drought crises in the mid-to-late 1990s under similar levels of low streamflow. This is attributed to water supply infrastructure investments planned before the late 1980’s drought (Fig. \ref{fig:Fig5}c shift from $Y_1$ to $Y_2$). The 2012 drought, while short in duration, led to crisis because it coincided with, and increased, peak summer water demand. Both the utility, whose ownership had been transferred to a not-for-profit public trust, and the state responded promptly, with prior state investment in water shortage planning likely facilitating this response. In the period between these two drought crises, attention and resources focused on wastewater treatment and water quality due to an EPA consent decree. While this reduced the resources available to focus on water supply robustness to drought, it may have also increased the utility’s capacity to address complex problems quickly (Fig. \ref{fig:Fig5}c). 

\section{Discussion}

We identified four biophysical archetypes (Table \ref{tab:Tab2}) that characterize the complexity of the historic biophysical environment (hydrology and infrastructure), and four institutional archetypes (Table 3) that characterize two key aspects of political-economic feedback control (information processing and response diversity). Configurations of these archetypes that are likely to contribute to dynamic fitness exhibit an institutional archetype capable of coping with biophysical complexity equal or greater than that observed (Fig. \ref{fig:Fig3}a). Alternate institutional designs result in different levels of resource investment in feedback capacity (Fig \ref{fig:Fig3}d, size of area within resource budget) and different divisions of resources between information processing capacity and response diversity (Fig. \ref{fig:Fig3}d, shape of boundary). Institutional designs that cope effectively with observed biophysical complexity while minimizing resources spent are efficient but have little buffer to cope with changing conditions. 

In addition to the total resources dedicated to feedback capacity, the proportion allocated to information processing and response diversity shapes dynamic fitness. When utilities invest in information processing, they can expand the range of conditions that they perceive to be well ordered and/or build capacity for exploratory information processing that enables effective response under conditions perceived as complex (Fig. \ref{fig:Fig3}b). Sufficient information processing capacity enables utilities to separate signal from noise, recognize when conditions change, and identify effective response actions. However, for a given institutional design to support fitness, it must also enable sufficient response diversity. For example, a utility with a constrained institutional structure may be able to identify the appropriate response but have insufficient response diversity to implement the response. In contrast, a utility with a decentralized institutional structure has high potential response diversity as authority to set goals and select actions is delegated to each local utility, but fragmentation limits the ability to share information and experiences and lack of higher level (e.g., state) planning requirements reduces incentives for exploratory information processing. If the biophysical complexity increases, a decentralized utility lacking the exploratory information processing to make sense of novel conditions may resort to trial-and-error (Fig \ref{fig:Fig3}c) or a reactive (wait-and-see) strategy. 

When we examine the patterns across the 16 cities, we do not find evidence of consistent application of institutional archetypes best fit for the biophysical archetypes present. While the moderate number of cases precludes broad generalization, we hypothesize that we see diversity in these pairings in part because different logics are applied in institutional design \citep{FrancoTorres2021} and because path dependency can override logics applied. For example, decentralization may be prioritized as a goal over efficiency, resulting in cases where the management of simple biophysical systems is delegated to local authorities, although centralization would be efficient \citep{Estache1995, Peterson2005}. In terms of path dependency, complex infrastructure may become highly constrained over time as multiple entities layer on regulation \citep{Ekanem2025, Lawless2024}. Further we acknowledge that while we characterized cases solely based on the biophysical complexity, there are other sources of complexity (e.g., diversity of demands and water users) that also drive institutional design. 

In examining patterns between archetype pairings and the timing, effectiveness, and response, we observe a mix of expected outcomes, informative contradictions and varying responses over time. Given the archetype pairing, the expected local level responsiveness is consistently observed in several cases across the study period (e.g., Atlanta, Boston, Phoenix, San Jose, Santa Rosa). Similarly, response was simply not needed in other cases where ample water availability and infrastructure fully addressed observed variability (e.g., Detroit, Hartford, Jacksonville, Memphis, Providence, Toledo, Washington). Variations in response across cities with common archetypes give us insight into the limitations of our analysis. For instance, San Jose and San Antonio were both classified into the dry with complex infrastructure and constrained archetypes but responded differently to drought crises. In San Jose there was strong state response and minimal city-level response across multiple droughts, as would be expected under a constrained institutional design. However, in San Antonio, there was a prompt and effective city-level response which we hypothesize to be influenced by the utility’s participation in an aquifer-wide planning process. San Antonio demonstrates that enabling conditions for dynamic fitness can be achieved through multiple institutional design mechanisms not all captured by our sorting criteria. 

Additionally, other factors influence response characteristics such as the level of government that leads in drought response. For example, when large areas of a state are affected by drought, state-level response may be needed to coordinate action amongst multiple water users withdrawing from the same sources (e.g., Harrisburg). Alternatively, a local water supplier may defer to a state-level response (e.g., order to conserve water during drought) to avoid spending their political capital on a repetitive or overlapping set of response actions (e.g., San Jose). This is consistent with polycentric systems enabling both competition and coordination to induce efficient responses across the network \citep{Aligica2012}. Further, we see patterns in the types of response actions taken, with the majority focused on information gathering,  planning, and water conservation. Infrastructure both creates and constrains opportunities for response. The high inertia of centralized water supply infrastructure results in high costs of augmenting or altering such infrastructure and the selection of other response actions \citep{Helmrich2023}. This high inertia highlights the importance of anticipatory capacity, as lead time is needed to implement infrastructure changes. 

The narratives of Charlotte, Harrisburg, and Indianapolis show how infrastructure and institutions co-evolve over time. In Charlotte (easy hydrology, simple infrastructure, enabled), we first observed a state-only response to drought in the early 2000s, specifically investments in information infrastructure. When subsequent droughts occurred in 2008 and 2012, the local response was prompt. Charlotte demonstrates dynamic fitness across multiple levels of governance as investments are made to enhance anticipatory and response capacity as needed (Fig \ref{fig:Fig5}a). In Harrisburg (seasonal hydrology, simple infrastructure, decentralized), we observe a prompt utility response to the first and second but not the third drought experienced in the study period. The third drought occurred during a period of financial stress and competing investment needs, highlighting the importance of resource availability (financial and otherwise) to translate responsiveness into response (Fig \ref{fig:Fig5}b). In Indianapolis (seasonal hydrology, complex infrastructure, constrained), response to the 1988 drought was delayed and occurred at only the state level, while the response to the 2012 drought was prompt and spanned local-utility-state levels in 2012. This pattern of response is hypothesized to be shaped both by a shift from a private to a not-for-profit trust water utility and capacity gained through response to other utility challenges (e.g., water quality), illustrating that institutional changes that do not shift the archetype may still change dynamic fitness (Fig \ref{fig:Fig5}c). 

The patterns in archetypes and response across cities and over time provides insight into the opportunities and tradeoffs in designing for dynamic fitness. The existing infrastructure and hydrological conditions shape the variability that the institutions must manage through feedback processes (Fig. \ref{fig:Fig1}). Institutional configurations create, enable or constrain these feedback pathways across a multilevel system (Fig. \ref{fig:Fig2}). From the institutional archetype development and application, we find that the distribution of control and flexibility across levels of governance shapes the information processing capacity and response diversity of utilities as well as the institutional costs. Configurations that optimize for efficiency under current conditions limit incipient capacity and are vulnerable to environmental change. Designs with buffer capacity in information processing and response diversity have higher costs. However, connectivity and coordination between utilities and state entities in the region can lower the costs of that additional capacity through sharing information and experience. Further, balance of information capacity and response diversity matters, and there are diminishing returns to investing in one without the other. In short, while institutional designs that enhance information processing capacity and response diversity at the utility level enable dynamic fitness at that same level, there are tradeoffs. Centralized systems maximize state (or more generally higher level) control and flexibility, and there are conditions under which a higher level entity such as the state government can be highly efficient at response (e.g., statewide CA droughts). Therefore, it is worth considering the scales of hazards a given infrastructure provider may face when considering tradeoffs in institutional design and balancing dynamic fitness across governance levels to address connectivity across spatial scales \citep{Moore2024}. There are also limits to the amount of variability that can be managed through institutional means (e.g., Hurricane Helene \citep{Cooper2024}), and there are conditions where preservation of performance will require infrastructure change. A minority of response actions observed resulted in new or altered infrastructure due to the high inertia of water supply infrastructure. This highlights the importance of anticipatory capacity to enable sufficient foresight to begin infrastructure investments early enough to allow for ample implementation time. 

In developing and testing archetypes we rely on historical data. This reliance on historical data presents a limitation as dynamic fitness is, by definition, forward looking as it gauges an institutional design's capacity to anticipate and respond to as of yet unexperienced future challenges. To address this limitation, we focus on historic events characterized as crises as they are conditions in which utilities are unable to solely rely on routine operations and have to strategize or react to less familiar conditions. The findings from this empirical analysis provide the initial linkages between broad theoretical advances to the urban water supply context and provide the foundation for hypothesizing on how to proactively design for dynamic fitness under diverse conditions.

\section{Conclusions}

As infrastructure systems increasingly operate outside their design conditions, attention is turning to institutional design to weather emerging challenges \citep{Chester2021, Moore2024, Wiechman2024}. While there is broad consensus that institutions must be fit to context to be successful \citep{Cox2012, Ostrom1995, Young2002}, a static definition of institutional fit is no longer appropriate in this era of rapid change \citep{Steffen2015} where the ability to dynamically shift between policies as conditions change shapes long term performance \citep{Anderies2019}. The essential challenge is that it is no longer sufficient for institutions to be fit (manage a stable variation regime), and instead they must be dynamically fit (capable of tracking environmental change). Dynamic fitness defined by responsiveness and anticipatory capacity is critical in the current era \citep{Moore2024}. To understand the features of institutional design that enable dynamic fitness and the biophysical characteristics that they are contingent upon, we focus on urban water supply systems as they epitomize the pressures of environmental and social change on infrastructure systems \citep{Krueger2022, Rockström2014}.  

Specifically, we draw on broad 35-year data sets for 16 U.S. urban water utilities to develop and test archetypes, which depict essential features of institutional design and the biophysical environment that lead to dynamic fitness. We hypothesize that configurations of these archetypes that are likely to contribute to dynamic fitness pair an institutional archetype capable of coping with biophysical complexity equal or greater than that observed. Institutional designs that develop both information processing capacity and response diversity are better able to handle more complex biophysical environments, though they incur higher institutional costs. However, these higher costs can be mitigated to a degree by polycentric governance at the regional level which facilitates sharing information and experiences. In contrast, institutional designs that optimize for efficiency under current environmental conditions save costs at the expense of vulnerability to environmental change. 

Further, there are tradeoffs in institutional design across levels of governance as the institutional structures that facilitate effective information processing and responsiveness at the utility level, may reduce the control and flexibility of higher levels of governance such as the state. Given the range of scales of change in the biophysical environment and the connectivity across scales, these tradesoffs deserve attention \citep{Moore2024}. Lastly, we observe few instances of changes to infrastructure in our case set likely because of the high inertia of infrastructure systems. However, it remains an important way to maintain performance and reduce the demands on institutions, and reinforces the need for anticipatory capacity to allow time for slow response actions such as infrastructure construction. 

\section*{Acknowledgements}

  All authors acknowledge support from the National Science Foundation Grant ``Transition Dynamics in Integrated Urban Water Systems'' (No. 1923880). The views and conclusions expressed here are those of the authors and do not necessarily reflect the views of the National Science Foundation.



\begin{appendix}

\subsection*{A: Case Characteristics} 

\begin{table*}[!h]
  \centering
     \caption{Summary of Case Characteristics}
    \begin{tabular}{cllcc}
    \toprule
       \textbf{City}  & Utility & Utility Type & \% Surface Water &  \% Groundwater  \\
       \midrule
        Atlanta, GA & Department of Watershed Management & municipal water utility & 100 & 0 \\
        Boston, MA & Boston Water and Sewer Commission & municipal water utility & 100 & 0\\ 
        Charlotte, NC & Charlotte Water & municipal water utility & 100 & 0 \\
        Detroit, MI & Detroit Water and Sewerage Department & municipal water utility & 100 & 0 \\
        Harrisburg, PA & Capital Region Water & special purpose unit & 100 & 0 \\
         & & of local government & & \\
        Hartford, CT & Metropolitan District Commission & municipal corporation & 100 & 0 \\
        Indianapolis, IN & Citizens Energy & public trust & 84 & 16 \\
        Jacksonville, FL & Jacksonville Electric Authority & community-owned utility & 0 & 100 \\
        Memphis, TN & Memphis Light and Gas and and Water & municipal water utility & 0 & 100 \\
        Phoenix, AZ & Phoenix Water Services Department & municipal water utility & 97 & 3 \\
        Providence, RI & Providence Water & municipal water utility, & 100 & 0 \\
         & & self-funded  & & \\
        Sacramento, CA & Department of Utilities’ Water Division & municipal water utility & 80 & 20 \\
        San Antonio, TX & Investor-owned public utility & 90 & 10 \\
        San Jose, CA & San Jose Water & investor-owned public utility & 50 & 50 \\
        Santa Rosa, CA & Water Department & municipal water utility & 93 & 7 \\
        Toledo, OH & Department of Public Utilities & municipal water utility & 100 & 0 \\
        Washington, DC & DC Water & independent authority of & 100 & 0 \\
         & & district government & & \\
    \bottomrule 
    \end{tabular}
\label{tab:Tab5}
\end{table*}

\subsection*{B: Response Coding Method} 
Two crisis events that occurred in 1988 (droughts in Boston and Indianapolis) were added to the dataset as there were clear responses in the study period. Events were coded by event category (authority, infrastructure, water source, financing, rates, policy, conservation and education) and governance level (utility, local, state or national). Across the sixteen cities this results in 488 discrete events, including 32 crisis events. One city, Providence, RI, had no identified crises.\\

The next step in the data analysis was to review the 10 years of events following each crisis event to identify and code response actions. Each crisis event response was coded in terms of timeliness, category of response, governance level(s) and effectiveness. Three categories of timeliness were defined: prompt with response in 0-2 years, medium 3-5 years, and delayed 5-10 years. If no response actions within 10 years were identified, the crisis event was coded as no response. The event categories and governance level described above were used and in both cases all applicable codes were noted (e.g., planning and infrastructure, local and state level). The effectiveness of the response was evaluated based on the presence or absence of a subsequent crisis triggered by a similar type and magnitude of disturbance within the following 10 years. If less than 10 years has passed since the crisis event the effectiveness is coded as unknown. 

\subsection*{C: Case Narratives} 

\subsubsection*{Indianapolis}

\textbf{Water Supply and Management in Indianapolis}\\
The Indianapolis metropolitan area water supply, currently operated by Citizens Energy Group, serves over 800,000 customers across the metropolitan region \citep{Citizens2025a}. Approximately, 60\% of the water supply comes from White River with an additional 15\% from Fall Creek, 9\% from Eagle Creek, 12\% from groundwater, and 4\% purchased from neighboring suppliers \citep{Citizens2013}. While located in a humid region with ample water supply under normal conditions, Indianapolis periodically experiences seasonal drought which can impact water supply operations. At the state level, there are no mandatory drought or conservation planning requirements in place in Indiana. However, Indiana has developed and implemented a voluntary water conservation and efficiency program (i.e., provides conservation planning frameworks for utilities) \citep{IDNR2025}. 
\\

\begin{figure*}[!h]
    \centering \includegraphics[width=0.6\linewidth]{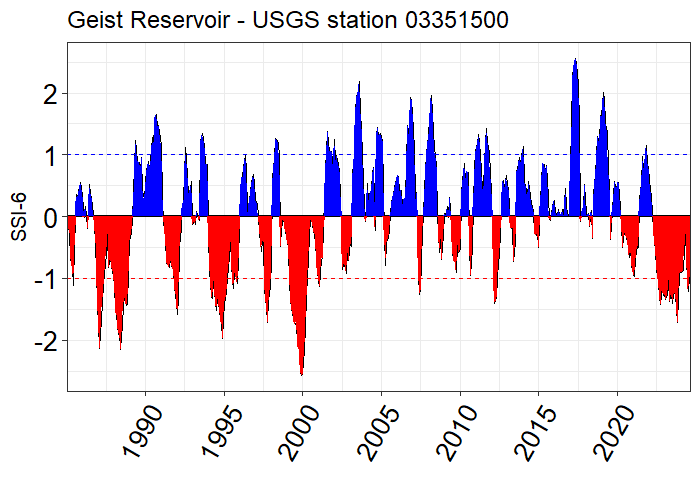}
    \caption{The Standardized Streamflow Index, SSI, provides a measure of the deviation from normal streamflow in a particular location over a particular averaging window \citep{Modarres2007}. Here we average over a 6-month window and show the SSI for the Geist Reservoir watershed for the full period from 1985 through the end of 2024. Red fill shows the below average periods while blue fill shows the above average periods. Values below -1 are considered drought.}
    \label{fig:S1}
\end{figure*}

\textbf{1988 Drought Crisis and Response}\\
Below average precipitation and above average temperatures in spring and summer of 1988 across the state of Indiana, led to drought across the state including in the supply watersheds for the city of Indianapolis \citep{USGS1992}. Fig. \ref{fig:S1} shows the streamflow into the Geist Reservoir, a key supply reservoir for Citizen’s Energy, is in drought as measured by the Standardized Streamflow Index (SSI). The drought impacted municipal water supply both directly through decreased streamflow and groundwater levels, and indirectly through increased agricultural water demand \citep{USDA1988, USGS1992}. 

While there was no documented city level response, the 1988 drought did prompt state level action. In 1991, the Indiana General Assembly passed HB 1260 which required the Indiana Department of Water Resources to develop a Water Shortage Plan \citep{Unterreiner2011}. The first Water Shortage Plan for the state was completed in 1994. Between 1989-1991, before completion of the plan, the Indianapolis Water Company (IWC, city water supplier) constructed three new water treatment plants, treating water from well fields and the White River. While these plants were in the works prior to the drought, they did provide additional capacity which increased robustness to droughts. A state Water Shortage Task Force was created by Indiana Code in 2006. The task force reviewed and updated the Water Shortage Plan and developed other resources for the state such as the Indiana Suggested Model Ordinance \citep{Unterreiner2011, IDNR2007}. The updated state Water Shortage Plan was published in 2009.
\\

\textbf{Management Changes and Compliance Challenges}\\
In 1997, the Northern Indiana Public Service Company (NIPSCO) acquired the IWC. However, the Indianapolis Water Company was the agency still managing the utility. NIPSCO was ordered to divest from IWC in 2000, and in 2002, the City of Indianapolis purchased the utility from the IWC.

In 2006, the EPA filed suit against the City of Indianapolis for violating the Clean Water Act due to combined sewer overflows. This consent decree was amended twice in 2009 and 2010, modifying almost half of the combined sewer overflow measures \citep{Citizens2025b}. While not directly affecting drought response or robustness, compliance with the consent degree required substantial staff and financial resources reducing capacity to address other issues. 
In 2011, the city sold the water and wastewater utilities to Citizens Energy Group (CEG), a non-profit public trust. In 2012, CEG began the DigIndy Tunnel System as the solution to reducing combined sewer overflows and remaining in compliance with the consent decree, which was amended a third time to describe the transfer of utility from the city to CEG. Citizens Water expanded its service area to Westfield in 2014 and constructed the Harbour Water Treatment Plant to treat water from the Harbour Well Field. Since CEG took over the operation of the water supply system there have been multiple water and sewer rate increases (e.g., 2014, 2016, 2017, and 2019); expenses related to the Consent Decree compliance were listed as the primary driver of the rate increases \citep{IndianaOffice2025}. 
2012 Drought Crisis and Response

Below average precipitation and above average temperatures resulted in drought for the Indianapolis water supply system in Spring and Summer of 2012 \citep{Citizens2013, NOAA2012}. In particular, in June 2012 precipitation was the lowest ever observed at the Indianapolis airport and the lowest SSI since 2000 (Fig. \ref{fig:S1}). Due to the extreme summer heat, water demand increased to a record high during June and July of 2012 \citep{Ballard2012}. In response to falling reservoir levels, the city introduced first voluntary and then mandatory conservation measures resulting in a temporary reduction of water demand by one third \citep{Ballard2012}. Demand reductions along with the return of rainfall led to the end of the drought water crisis.

After the immediate drought crisis abated, both the CEG and the state of Indiana responded with measures to decrease the risk of water supply impacts from a future drought. CEG published its own Drought Management Plan in 2013 \citep{Citizens2013} and in 2015 it announced the construction of a new reservoir to increase water storage \citep{Delany2015}. This reservoir, named Citizens Reservoir, came on line in 2020 \citep{Citizens2025c}. In early 2012 before the drought was evident, the State legislature repealed the previously established State Water Shortage Taskforce, so the task force had no role in drought response. However, the State of Indiana released another update to the state Water Shortage Plan in 2015 \citep{IDNR2015}.

\subsubsection*{Charlotte}

\textbf{Water Supply and Management in Charlotte} \\
Charlotte Water is the largest public water and wastewater utility in North Carolina, serving over one million customers in the City of Charlotte and throughout greater Mecklenburg County \citep{CharlotteWater2025a}. This service area includes the city of Charlotte and the towns of Matthews, Mint Hill, Pineville, Huntersville, Davidson, and Cornelius. Charlotte Water is a public utility operated by the City of Charlotte and has been in operation since 1899 \citep{McNeely2023}. The main water source is the Catawba-Wateree River Basin with intakes at Lake Norman and Mountain Island Lake \citep{CharlotteWater2025b}. Charlotte Water It is regulated by the North Carolina Environmental Management Commission and the North Carolina Department of Environmental Quality.

\begin{figure*}[h]
    \centering \includegraphics[width=0.6\linewidth]{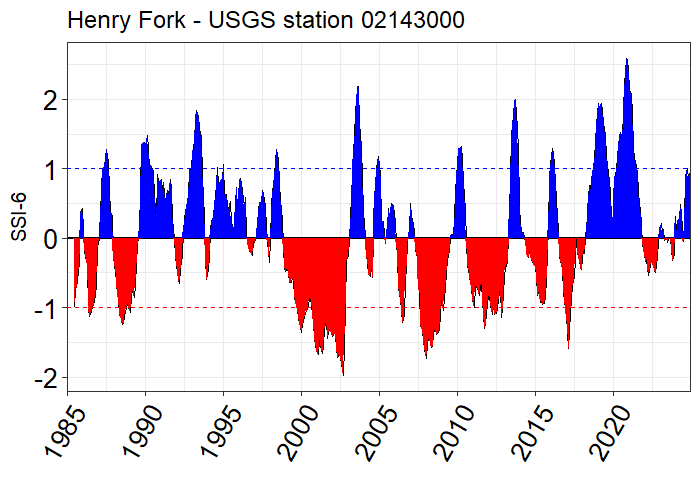}
    \caption{The SSI with a 6-month averaging window for the Henry Fork River which is upstream of the Norman Reservoir for the period from 1985 through the end of 2024. Red fill shows the below average periods while blue fill shows the above average periods. Values below -1 are considered drought.}
    \label{fig:S2}
\end{figure*}

While relatively wet (temperate climate with no dry season, Peel et al. 2007) Charlotte still has to contend with drought and faces the challenges of increasing demand and rainfall variability. While recent years have been wetter than average (Fig. \ref{fig:S2}, 2018-2022), historically, Charlotte has dealt with several droughts. For example, the crippling 1911 event that required water to be brought in via rail and provided impetus to build a more reliable water pumping system from the Catawba River and create a much larger reservoir at Mountain Island Lake \citep{CharlotteWater2023} with three recent extended drought periods - 2002, 2008, and 2012 that are readily apparent in the SSI in Fig. \ref{fig:S2}.  
\\

\textbf{2002 Drought Crisis and Response}

Precipitation deficits during the 1998–2002 drought for some locations in North Carolina were among the largest documented since the beginning of systematic collection of weather data.  This 4-year drought period caused widespread hardship and economic losses and toward the end of 2002, more than 200 municipalities operated under some form of voluntary, mandatory, or emergency water conservation. Reservoirs across North Carolina were at record or near record-low levels and required careful operation to balance the needs of upstream and downstream users \citep{USGS2005}. 

In response, the General Assembly passed Session Law 2002-167, later codified as General Statute 143-355.1 giving statutory authority to the Drought Management Advisory Council (DMAC) to provide objective drought status advisories based on technical data and are to be crafted to fit varying conditions in different parts of the state \citep{DMAC2025}. The legislation further compelled the Environmental Management Commission to develop and implement rules for water supply planning, conservation, and reuse during droughts. Soon after, the 2006 Water Supply Study completed for relicensing of Duke Energy’s Catawba-Wateree Hydroelectric Project included an evaluation of future water use projections for the Basin through the year 2058 that highlighted potential challenges for the sustainability of the regions water supply. This, in turn, led to the formation of the Catawba-Wateree Water Management Group (CWWMG) in 2006. The CWWMG is a non‐profit corporation tasked with identifying, funding, and managing projects to balance water resources for human needs while maintaining the river’s ecological health by developing a water supply master plan \citep{CWWMG2014}. One outcome was the development of the Low Inflow protocol (LIP) to establish procedures for reductions in water use during periods of low inflow to the Catawba-Wateree basin in 2006 implemented by the CW Drought Management Advisory Group (DMAG) \citep{CWWMG2014}.
\\

\textbf{2008 and 2012 Drought Crises and Responses}

Droughts in 2008 and 2012 continued to drive the development of drought management investments building on the CWWMG, the CWDMAG, and the LIP. Developments included the Smart Irrigation Program which incentivised customers to equip their irrigation systems with approved state of the art equipment to reduce water demand and curtail water runoff \citep{WBTV2011, CharlotteWater2025c} and building on the Envision Charlotte’s energy program who worked with local stakeholders to address the complex issue of sustainable water resources. Charlotte Water joined the project to address water usage and delivery, aiming to ultimately reach the goal of a significant reduction in water usage (Envision Charlotte, 2018). Charlotte-Mecklenburg Utilities continued to participate in collaborative research regarding sustainable water management. For example, the 2013 Defining and Enhancing the Safe Yield of a Multi-Use, Multi-Reservoir Water Supply study survey which, when combined with the southeastern U.S. study helps define strategies to assure safe yields of water supply \citep{WRF2013}.  The CWWMG’s mandate to produce a water supply master plan was realized in 2014. The plan focuses on protecting, preserving, and extending the available water supply in the Catawba-Wateree River and its 11 reservoirs. 
\\

\textbf{Summary of Institutional Dynamics}

Charlotte water remains very active with conservation efforts and focuses on minimizing water use and waste \citep{Frost2025}. The LIP and Water Shortage management plan (written in 2003, subsequently updated 5 times) remain central to managing drought conditions. The LIP provides a tiered (5 stages from “watch” to “emergency”) feedback control mechanism to reduce water use to match availability. Recent innovations on the conservation space include Charlotte water partnering with xylem, a water technology company, and a local brewery, Town Brewing, to brew beer (“Renew Brew”) using polished wastewater treatment plant effluent \citep{QCWater2025}. CLTWater has a robust capital investment plan as well. 
\\

\subsubsection*{Harrisburg}

\textbf{Water Supply and Management in Harrisburg}\\
The Harrisburg metropolitan area water supply, currently operated by Capital Region Water, serves over 60,000 residents and businesses in the City of Harrisburg and municipalities in Dauphin County, including parts of the Penbrook Borough and four neighboring townships \citep{CRW2025a}. The primary water supply comes from the DeHart Reservoir, about 20 miles from Harrisburg, fed by Clark Creek, and the Susquehanna River serves as a secondary source available in emergencies \citep{CRW2025b}. Harrisburg is located in a humid region with ample water supply under normal conditions, but periodically experiences seasonal drought, which can impact water supply availability. At the state level, there are no mandatory drought or conservation planning requirements; however, in accordance with regulation adopted in 1981 and amended in 1991 and 2001, Pennsylvania requires water utilities to implement water conservation measures if the governor issues a proclamation declaring a drought-related state of emergency.
\\

\begin{figure*}[h]
    \centering \includegraphics[width=0.6\linewidth]{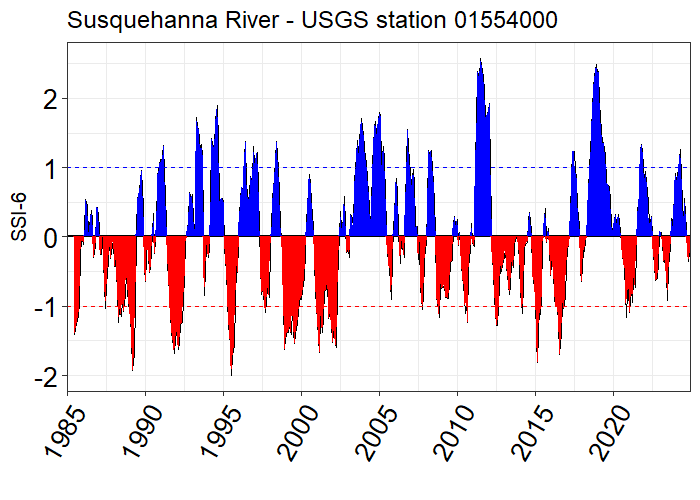}
    \caption{The SSI with a 6-month averaging window for the SSI for the Susquehanna River which for the period from 1985 through the end of 2024. Red fill shows the below average periods while blue fill shows the above average periods. Values below -1 are considered drought. After the 1991 drought, two new infrastructure projects came online in 1994, including the Susquehanna River Pump Station and the Dr. Robert E. Young Water Services Center (a water filtration plant). The pump station was built to source water from the Susquehanna River when the flows of the DeHart Reservoir are low \citep{CRW2021a}. Both initiatives collectively increased the city’s water management and distribution capacity.}
    \label{fig:S3}
\end{figure*}

\textbf{Drought Crises and Responses}\\
Since 1990, three statewide drought emergencies were declared in Pennsylvania and directly impacted Dauphin County, including between July 1991 and April 1992, July to September 1999, and February to November 2002 \citep{PDEP2025}. Fig. \ref{fig:S3} shows the streamflow into the Susquehanna River, the river flowing through Harrisburg, which serves as an emergency source, was in drought as measured by the Standardized Streamflow Index (SSI) during each of the aforementioned drought emergency time periods. During each period, there were prolonged periods of decreased flows. Each emergency drought declaration from the governor’s office called on residents to conserve water and required water utilities operating in counties with emergency drought declarations to take actions aimed at conserving water. These mandatory requirements primarily include reducing outdoor watering (e.g., watering grass, athletic fields, outdoor gardens, golf courses, washing paved surfaces, washing cars, etc.) \citep{Penn}. The following paragraphs discuss specific actions taken after each drought period.
\\

\textbf{Management Changes and Compliance Challenges}\\
Although the city's last emergency drought declaration occurred in 2002, ongoing compliance and management challenges have led to multiple local initiatives that have improved the system’s robustness to drought and other environmental challenges. In August 2013, the city adopted the Harrisburg Strong Plan, which outlined steps for fiscal recovery, including a major restructuring of the city’s water, stormwater, and sewer systems \citep{Harrisburg2013}. In late 2013, operational control and ownership of these systems were transferred to Capital Region Water (CRW) – an independent municipal authority – enabling it to independently manage infrastructure upgrades, comply with environmental regulations, and regain access to capital markets \citep{Raftelis2025}.
 
After acquiring control, CRW took multiple actions aimed at improving management and infrastructure upgrades, including the adoption of a geographic information system (GIS) asset management program in 2013 \citep{GeoDecisions2024} and a water rate increase in 2014 \citep{CRW2014}. In 2015, CRW entered into a Partial Consent Decree with the EPA and the Pennsylvania Department of Environmental Protection (PDEP) to address violations of the Clean Water Act and the Pennsylvania Clean Streams Law \citep{US2015}. The decree required CRW to reduce combined sewer overflows, eliminate severe sewer system infrastructural defects, and develop a Long-Term Control Plan that includes consideration of green infrastructure solutions. 

CRW subsequently began a 2015 source water protection plan for DeHart Reservoir and Susquehanna River \citep{CRW2015a} – referred to as City Beautiful H20 \citep{CRW2025d} – and took multiple actions that align with the decree’s requirements, including the construction of a new advanced wastewater treatment facility \citep{CRW2015b}, approval of an agreement to conserve its 8,200 acre DeHart property by easement in 2016 \citep{CRW2016}, and adoption of a community greening plan in 2017 (plan focused on improving urban stormwater infrastructure) \citep{CRW2017}. The City Beautiful H20 planning was completed in 2018, and seven fee and rate increases (water, sewer, and stormwater) were adopted from 2018 through 2024 to fund new infrastructure upgrades \citep{Binda2018, CRW2024, CRW2020a, CRW2020b, CRW2021b, Miller2017, Raftelis2019}. In 2024, CRW adopted a plan to update City Beautiful H20 to further expand infrastructure robustness in the community \citep{CRW2024}, and a new initiative to update water meters is underway \citep{CRW2025c}.

    \end{appendix}


\begin{thebibliography}{}

\bibitem[Aggarwal and Anderies, 2023]{Aggarwal2023}
Aggarwal, R. and Anderies, J.~M. (2023).
\newblock Understanding how governance emerges in social-ecological systems: insights from archetype analysis.
\newblock {\em Ecology and Society}, 28(2).
\newblock \_eprint: https://doi.org/10.5751/es-14061-280202.

\bibitem[Aligica and Tarko, 2012]{Aligica2012}
Aligica, P.~D. and Tarko, V. (2012).
\newblock Managing variance: {Key} policy challenges for the {Anthropocene}.
\newblock {\em Governance}, (Oxford, England), 25, 237–262.

\bibitem[Alonso Vicario et al., 2024]{Alonso2024}
Alonso Vicario, S., Hornberger, G.~M., Mazzoleni, M., and Garcia, M.  (2024).
\newblock The importance of climate and anthropogenic influence in precipitation partitioning in the contiguous {United} {States}.
\newblock {\em Journal of Hydrology}, 633(130984), 130984.

\bibitem[Anderies, 2015]{Anderies2015}
Anderies, J.~M. (2015).
\newblock Managing variance: {Key} policy challenges for the {Anthropocene}.
\newblock {\em Proceedings of the National Academy of Sciences of the United
  States of America}, 112(47):14402--14403.

\bibitem[Anderies and Janssen, 2013]{Anderies2013}
Anderies, J.~M. and Janssen, M.~A. (2013).
\newblock Robustness of social-ecological systems: {Implications} for public
  policy.
\newblock {\em Policy Studies Journal}, 41(3):513--536.

\bibitem[Anderies et~al., 2019]{Anderies2019}
Anderies, J.~M., Mathias, J.~D., and Janssen, M.~A. (2019).
\newblock Knowledge infrastructure and safe operating spaces in
  social–ecological systems.
\newblock {\em Proceedings of the National Academy of Sciences of the United
  States of America}, 116(12):5277--5284.

\bibitem[Ashby, 1968]{Ashby1968}
Ashby, W.~R.  (1968).
\newblock  Principles of the Self-Organizing System.
\newblock In W. Buckley (Ed.), {\em Systems Research for Behavioral Science}, Routledge.

\bibitem[Azizi et~al., 2024]{Azizi2024}
Azizi, K., Hornberger, G.~M., Baggio, J., Koebele, E.~A., Anderies, J.~M., and Garcia, M.  (2024).
\newblock Conditions that Support the Provision of High-Quality and Affordable Urban Drinking Water in the US.
\newblock {\em Journal of Water Resources Planning and Management}, 150.
\newblock \_eprint: https://doi.org/10.1061/JWRMD5.WRENG-6289.

\bibitem[Baggio et al., 2016]{Baggio2016}
Baggio, J.~A., Barnett, A.~J., Perez-Ibarra, I., Brady, U., Ratajczyk, E., Rollins, N., Rubiños, C., Shin, H.~C., Yu, D. J., Aggarwal, R., Anderies, J.~M., and Janssen, M.~A. (2016).
\newblock Explaining success and failure in the commons: the configural nature of {Ostrom’s} institutional design principles.
\newblock {\em International Journal of the Commons}, 10(2), 417.

\bibitem[Ballard, 2012]{Ballard2012}
Ballard, G.~A. (2012).
\newblock Gregory A. Ballard, Mayor of Indianapolis, Testimony Before the U.S. House of Representatives
\newblock {\em Committee On Science, Space and Technology}.
\newblock Accessed on 1/31/2025: https://democrats-science.house.gov/imo/media/doc/Ballard \%20Testimony.pdf

\bibitem[Binda, 2018]{Binda2018}
Binda, L. (2018).
\newblock Capital Region Water sets 2019 budget, will increase water, sewer rates. 
\newblock {\em The Burg News}.
\newblock Accessed on 1/31/2025: https://theburgnews.com/news/crw-sets-2019-budget-will-increase-water-sewer-rates

\bibitem[Boisot and Mckelvey, 2011]{Boisot2011}
Boisot, M., and Mckelvey, B.   (2011).
\newblock  Complexity and Organization–Environment Relations: Revisiting Ashby’s Law of Requisite Variety.
\newblock In {\em The Sage Handbook of Complexity and Management}, SAGE Publications Ltd.

\bibitem[Carlson and Doyle, 2002]{Carlson2002}
Carlson, J.~M., and Doyle, J. (2002).
\newblock Complexity and robustness.
\newblock {\em Proceedings of the National Academy of Sciences of the United
  States of America}, 99(suppl 1):2538–2545.
  
\bibitem[Capital Region Water, 2025a]{CRW2025a}
Capital Region Water (2025a).
\newblock Water Quality Reports. 
\newblock Accessed on 2/19/2025: https://capitalregionwater.com/resources/water-quality-reports/

\bibitem[Capital Region Water, 2025b]{CRW2025b}
Capital Region Water (2025b).
\newblock Ensuring Clean Water. 
\newblock Accessed on 2/19/2025: https://capitalregionwater.com/what-we-do/ensuring-clean-water/\#:\~:text=Drinking\%20 Water,you\%20turn\%20on\%20the\%20faucet.

\bibitem[Capital Region Water, 2025c]{CRW2025c}
Capital Region Water (2025c).
\newblock Advanced Meter Infrastructure (AMI) Upgrade.
\newblock Accessed on 3/28/2025: https://capitalregionwater.com/projects/drinking-water-distribution-system-automated-meter-reading-network-conversion/

\bibitem[Capital Region Water, 2025d]{CRW2025d}
Capital Region Water (2025d).
\newblock City Beautiful H2O: Helping Harrisburg tackle runoff pollution
\newblock Accessed on 3/28/2025: https://capitalregionwater.com /what-we-do/cbh2o/

\bibitem[Capital Region Water, 2024]{CRW2024}
Capital Region Water (2024).
\newblock Alternative Analysis Report. 
\newblock Accessed on 2/19/2025: https://capitalregionwater.com/ wp-content/uploads/2024/04/ CRW\_Alternatives\_Analysis\_Report.pdf

\bibitem[Capital Region Water, 2021a]{CRW2021a}
Capital Region Water (2021a).
\newblock Consulting Engineer’s Annual Report – Water System. 
\newblock Accessed on 2/19/2025: https://capitalregionwater.com/wp-content/uploads/2021/11/2021-09-30-Water-CEAR-FINAL-2021.pdf

\bibitem[Capital Region Water, 2021b]{CRW2021b}
Capital Region Water (2021b).
\newblock Capital Region Water Approves 2022 Budgets and Rates. 
\newblock Accessed on 2/19/2025: https://capitalregionwater.com/news/capital-region-water-approves-2022-budgets-and-rates/

\bibitem[Capital Region Water, 2020a]{CRW2020a}
Capital Region Water (2020a).
\newblock Financial Statements and Required Supplementary Information and Supplementary Information. 
\newblock Accessed on 2/19/2025: https://capitalregionwater.com/wp-content/uploads/2021/09/CRW-Financial-Statements-2020.pdf

\bibitem[Capital Region Water, 2020b]{CRW2020b}
Capital Region Water (2020b).
\newblock Capital Region Water Approves 2021 Budgets and Rates. 
\newblock Accessed on 2/19/2025: https://capitalregionwater.com/news/capital-region-water-approves-2021-budgets-and-rates/\#:\~:text=Drinking\%20water\%20rates\%20will\%20increase \%20by\%202\%,charge\%20of\%20\$7.93\%20for \%20a\%20standard\%205/8\%E2\%80\%9D

\bibitem[Capital Region Water, 2017]{CRW2017}
Capital Region Water (2017).
\newblock Community Greening Plan: A Green Stormwater Infrastructure Plan for Harrisburg. 
\newblock Accessed on 1/15/2026: https://capitalregionwater.com/wp-content/uploads/2017/04/Community-Greening-Plan\_FINAL\_smallest.pdf

\bibitem[Capital Region Water, 2016]{CRW2016}
Capital Region Water (2016).
\newblock Historic Agreement Permanently Protects Source of Harrisburg’s Drinking Water. 
\newblock Accessed on 2/19/2025: https://capitalregionwater.com/wp-content/uploads/2014/05/Historic-Agreement-Protects-Drinking-Water1.pdf

\bibitem[Capital Region Water, 2015a]{CRW2015a}
Capital Region Water (2015a).
\newblock Source Water Protection Plan DeHart Reservoir and Susquehanna River. 
\newblock Accessed on 2/19/2025: https://capitalregionwater.com/wp-content/uploads/2016/02/2015-09-00-SWP-Final-Screen-v1.pdf

\bibitem[Capital Region Water, 2015b]{CRW2015b}
Capital Region Water (2015b).
\newblock Request for Proposals for Professional Engineering Services
Front Street Pump Station Upgrade ADDENDUM NO. 1.
\newblock Accessed on 2/19/2025: https://capitalregionwater.com/wp-content/uploads/2014/06/2015-05-27-Addendum-No.-1-to-FSPS-Upgrade-RFP.pdf\#page=1.31

\bibitem[Capital Region Water, 2014]{CRW2014}
Capital Region Water (2014).
\newblock Water Rates. 
\newblock Accessed on 2/19/2025: https://cityofharrisburg.zendesk.com/hc/en-us/article\_attachments/202266260

\bibitem[Charlotte Water, 2025a]{CharlotteWater2025a}
Charlotte Water (2025a).
\newblock About Us. 
\newblock Accessed on 4/30/2025: https://www.charlottenc.gov/water/About-Us.

\bibitem[Charlotte Water, 2025b]{CharlotteWater2025b}
Charlotte Water (2025b).
\newblock Charlotte Water IBT. 
\newblock Accessed on 4/30/2025: https://www.charlottenc.gov/water/ Water-Quality/Charlotte-Water-IBT.

\bibitem[Charlotte Water, 2025c]{CharlotteWater2025c}
Charlotte Water (2025c).
\newblock Smart Irrigation. 
\newblock Accessed on 4/30/2025: https://www.charlottenc.gov/water /Programs/Smart-Irrigation.

\bibitem[Charlotte Water, 2023]{CharlotteWater2023}
Charlotte Water (2023).
\newblock Charlotte Water Annual Report. 
\newblock Accessed on 4/30/2025: https://www.charlottenc.gov/files/
sharedassets/cltwater/v/1/documents/about-us/cltw \_annual-report\_2023\_fnl.pdf.

\bibitem[CWWMG, 2014]{CWWMG2014}
Catawba-Wateree Water Management Group [CWWMG] (2014).
\newblock Catawba-Wateree River Basin Water Supply Master Plan. 
\newblock Accessed on 4/30/2025: https://www.charlottenc.gov/files/
sharedassets/cltwater/v/1/documents/about-us/cltw \_annual-report\_2023\_fnl.pdf.

\bibitem[Chester et~al., 2021]{Chester2021}
Chester, M., Underwood, B.~S., Allenby, B., Garcia, M., Samaras, C., Markolf, S., Sanders, K., Preston, B., and Miller, T.~R. (2021).
\newblock Infrastructure resilience to navigate increasingly uncertain and complex conditions in the {Anthropocene}.
\newblock {\em Npj Urban Sustainability}, 1.
\newblock \_eprint: https://doi.org/10.1038/s42949-021-00016-y.

\bibitem[Chester and Allenby, 2021]{ChesterAllenby2021}
Chester, M. and Allenby, B. (2021).
\newblock Toward adaptive infrastructure: the Fifth Discipline.
\newblock {\em Sustainable and Resilient Infrastructure}, 6, 334–338.

\bibitem[Citizens Energy Group, 2013]{Citizens2013}
Citizens Energy Group (2013).
\newblock Drought Management Plan.
\newblock Accessed on 1/31/2025: https://info.citizensenergygroup.com/conservation/water/ drought-alerts 

\bibitem[Citizens Energy Group, 2025a]{Citizens2025a}
Citizens Energy Group (2025a).
\newblock Water.
\newblock Accessed on 1/31/2025: https://info.citizensenergygroup.com/water

\bibitem[Citizens Energy Group, 2025b]{Citizens2025b}
Citizens Energy Group (2025b).
\newblock Long Term Control Plan.
\newblock Accessed on 1/31/2025: https://info.citizensenergygroup.com/digindy/regulation/ control-plan 

\bibitem[Citizens Energy Group, 2025c]{Citizens2025c}
Citizens Energy Group (2025c).
\newblock  Citizens Reservoir: Water Supply for the Future.
\newblock Accessed on 1/31/2025: https://info.citizensenergygroup.com/projects/citizens-reservoir 

\bibitem[City of Harrisburg, 2013]{Harrisburg2013}
City of Harrisburg. (2013).
\newblock  Harrisburg Strong Plan.
\newblock Accessed on 1/31/2025: https://dced.pa.gov/download/harrisburg-strong-plan-pdf/

\bibitem[Commonwealth of Pennsylvania, n.d.]{Penn}
Commonwealth of Pennsylvania. (n.d.).
\newblock  Title 4, Chapter 119: Prohibition of nonessential water uses in a Commonwealth drought emergency area. Pennsylvania Code.  
\newblock Accessed on 1/31/2025: https://www.pacodeandbulletin.gov/ Display/pacode?file=/secure/pacode/data/004/ chapter119/chap119toc.html

\bibitem[Cooper, 2024]{Cooper2024}
Cooper, R. (2024).
\newblock Hurricane Helene Recovery.
\newblock Office of State Budget and Management.

\bibitem[Cox, 2012]{Cox2012}
Cox, M. (2012).
\newblock Diagnosing institutional fit: A formal perspective.
\newblock {\em Ecology and Society}, 17(4).
\newblock \_eprint: https://doi.org/10.5751/ES-05173-170454

\bibitem[Crawford and Ostrom, 1995]{Crawford1995}
Crawford, S.~E.~S., and Ostrom, E. (1995).
\newblock  A grammar of institutions.
\newblock {\em The American Political Science Review}, 189(3), 582–600.

\bibitem[Csete and Doyle, 2002]{Csete2002}
Csete, M.~E., and Doyle, J.~C. (2002).
\newblock Reverse engineering of biological complexity
\newblock {\em Science}, 295(5560), 1664–1669.

\bibitem[Daems and Schaefer, 2024]{Daems2024}
Daems, D., and Schaefer, A. (2024).
\newblock The problem of complexity and the emergence of polycentric political order.
\newblock  In D. Thunder and P. Paniagua (Eds.) {\em Polycentric governance and the good society: a normative and philosophical investigation}, (1st ed., pp. 87–113). Lexington.

\bibitem[Davis et~al., 2010]{Davis2010}
Davis, S. J., Caldeira, K., and Matthews, H. D.  (2010).
\newblock Future CO 2 Emissions and Climate Change from Existing Energy Infrastructure. 
\newblock {\em Science}, 329, 1330–1333.

\bibitem[DeCaro et~al., 2017]{DeCaro2017}
DeCaro, D.~A., Chaffin, B.~C., Schlager, E., Garmestani, A.~S., and Ruhl, J.~B.  (2017).
\newblock Legal and institutional foundations of adaptive environmental governance.
\newblock {\em Ecology and Society}, 22(32).

\bibitem[Delany, 2015]{Delany2015}
Delany, R. (2015).
\newblock  Citizens Will Add To Water Supply By Building 2.7 Billion Gallon Reservoir In Fishers.
\newblock Accessed on 2/12/2025: https://www.wfyi.org/news/articles/citizens-will-add-to-water-supply-by-building-27-billion-gallon-reservoir-in-fishers 

\bibitem[Deslatte et~al., 2025]{Deslatte2025}
Deslatte, A., Adams, J.~A., Cheema, F., Barnes, J.~L., Koebele, E.~A., and Alonso~Vicario, S. (2025).
\newblock Understanding the Impact of Institutions on Climate‐Adaptive Policy Designs: A Study of Collective Action Inference in Urban Water Systems.
\newblock {\em Policy Studies Journal}, 53, 637–653.

\bibitem[Deslatte et~al., 2024a]{DeslatteAdams2024}
Deslatte, A., Adams, J., Cheema, F., Alonso~Vicario, S., Barnes, J., and Koebele, E.~A.  (2024).
\newblock How state-reinforced knowledge infrastructure influences adaptive urban water governance.
\newblock {\em Ecology and Society}, 29(28).

\bibitem[Deslatte et~al., 2022]{Deslatte2022}
Deslatte, A., Helmke‐Long, L., Anderies, J.~M., Garcia, M., Hornberger,
  G.~M., and Koebele, E.~A. (2022).
\newblock Assessing sustainability through the {Institutional} {Grammar} of
  urban water systems.
\newblock {\em Policy Studies Journal}, 50:387--406.

\bibitem[Deslatte et~al., 2023]{Deslatte2023}
Deslatte, A., Koebele, E.~A., Bartels, L., Wiechman, A., Alonso~Vicario, S., Coughlin, C., and Rybolt, D. (2023).
\newblock Institutions, Voids, and Dependencies: Tracing the Designs and Robustness of Urban Water Systems.
\newblock {\em International Review of Public Policy}, 5, 180–222.

\bibitem[Deslatte et~al., 2024b]{DeslatteKoebele2024}
Deslatte, A., Koebele, E.~A., and Wiechman, A. (2024).
\newblock  Embracing the ambiguity: Tracing climate response diversity in urban water management.
\newblock {\em Public Administration}, 103(1), 250-272.
\newblock \_eprint: https://doi.org/10.1111/padm.13017

\bibitem[Dettinger and Diaz, 2000]{Dettinger2000}
Dettinger, M.~D., and Diaz, H.~F. (2000).
\newblock Global Characteristics of Stream Flow Seasonality and Variability.
\newblock {\em Journal of Hydrometeorology}, 1, 289-310.

\bibitem[Drought Management Advisory Council, 2025]{DMAC2025}
Drought Management Advisory Council (2025).
\newblock The DMAC Process.
\newblock Accessed on 1/31/2025: https://www.ncdrought.org/DMAC-text-new

\bibitem[Eisenack et~al., 2006]{Eisenack2006}
Eisenack, K., Luedeke, M., and Kropp, J. (2006).
\newblock Construction of archetypes as a formal method to analyze socialecological systems
\newblock {\em Proceedings of the Institutional Dimensions of Global Environmental Change}.

\bibitem[Eisenack et~al., 2021]{Eisenack2021}
Eisenack, K., Oberlack, C., and Sietz, D.  (2021).
\newblock Avenues of archetype analysis: roots, achievements, and next steps in sustainability research. 
\newblock {\em Ecology and Society}, 26(2). 

\bibitem[Ekanem et~al., 2025]{Ekanem2025}
Ekanem, M., Noble, B., and Poelzer, G.   (2025).
\newblock The effects of institutional layering on electricity sector reform: Lessons from Norway’s electricity sector. 
\newblock {\em Energy Research and Social Science}, 119, 103864. 

\bibitem[Ekstrom and Young, 2009]{Ekstrom2009}
Ekstrom, J.~A., and Young, O.~R. (2009).
\newblock Evaluating Functional Fit between a Set of Institutions and an Ecosystem.
\newblock {\em Ecology and Society}, 14(2). 

\bibitem[Envision Charlotte, 2018]{EnvisionCharlotte2018}
Envision Charlotte (2018).
\newblock Envision Charlotte: Water Report - Water Benchmarking and Conservation. 
\newblock Accessed on 6/29/2019: https://envisioncharlotte.com/wp-content/ uploads/2018/03/101594WP-01\_Envision-Charlotte-White-Paper-1.pdf

\bibitem[Estache and Sinha, 1995]{Estache1995}
Estache, S., and Sinha, A. (1995).
\newblock Does decentralization increase spending on public infrastructure?
\newblock {\em World Bank Publications}. No. 1457

\bibitem[Franco-Torres et al., 2021]{FrancoTorres2021}
Franco-Torres, M., Kvålshaugen, R., and Ugarelli, R.~M. (2021).
\newblock Understanding the governance of urban water services from an institutional logics perspective.
\newblock {\em Utilities Policy}, 68, 101159.

\bibitem[Frost, 2025]{Frost2025}
Frost, J. (2025).
\newblock Charlotte Water Conservation and Water Use Programs.
\newblock Accessed on 4/30/2025: https://www.youtube.com/watch?v=Hii2j03TYsc

\bibitem[Garcia et~al., 2025]{Garcia2025}
Garcia, M., Wiechman, A., Alonso Vicario, S., Hornberger, G.~M., Bartels, L., Azizi, K., Anderies, J.~M., Koebele, E.~A., Barnes, J., Deslatte, A., and Adams, J.  (2025).
\newblock U.S. Water Utility Timelines 1990-2020
\newblock {\em CUASHI Hydroshare}.
\newblock http://www.hydroshare.org/resource/ 7a1d900ea3c84676a417089ef4e0d7b9.

\bibitem[Garcia et al., 2019]{Garcia2019}
Garcia, M., Koebele, E.~A., Deslatte, A., Ernst, K., Manago, K.~F., and Treuer, G.  (2019).
\newblock Towards urban water sustainability: Analyzing management transitions in Miami, Las Vegas, and Los Angeles.
\newblock {\em Global Environmental Change}, 58(101967), 101967.

\bibitem[Garcia et al., 2020]{Garcia2020}
Garcia, M., Ridolfi, E., and Di Baldassarre, G.  (2020).
\newblock The interplay between reservoir storage and operating rules under evolving conditions.
\newblock {\em Journal of Hydrology}, 590(125270), 125270.

\bibitem[GeoDecisions, 2024]{GeoDecisions2024}
GeoDecisions. (2024).
\newblock Capital Region Water taps GeoDecisions for GIS and computerized maintenance management systems.
\newblock Accessed on 3/28/2025: https://www.geodecisions.com/news/2024/6/24/capital-region-water-taps-geodecisions-for-gis-and-computerized-maintenance-management-systems

\bibitem[Gilrein et al., 2019]{Gilrein2019}
Gilrein, E.~J., Carvalhaes, T.~M., Markolf, S.~A., Chester, M.~V., Allenby, B.~R., and Garcia, M. (2019).
\newblock Concepts and practices for transforming infrastructure from rigid to adaptable.
\newblock {\em Sustainable and Resilient Infrastructure}, 6(3-4), 213-234.

\bibitem[Gonzales and Ajami, 2017]{Gonzales2017}
Gonzales, P., and Ajami, N.~K. (2017).
\newblock Gonzales, P., and Ajami, N. K. (2017). An integrative regional resilience framework for the changing urban water paradigm.
\newblock {\em Sustainable Cities and Society}, 30, 128–138.

\bibitem[Güntner et~al., 2007]{Güntner2007}
Güntner, A., Stuck, J., Werth, S., Döll, P., Verzano, K., and Merz, B. (2024).
\newblock A global analysis of temporal and spatial variations in continental water storage.
\newblock {\em Water Resources Research}, 43(5).

\bibitem[Helmrich et~al., 2023]{Helmrich2023}
Helmrich, A., Chester, M., Miller, T.~R., and Allenby, B. (2023).
\newblock Lock-in: origination and significance within infrastructure systems.
\newblock {\em Environmental Research: Infrastructure and Sustainability},
  3(3):032001.
\newblock Publisher: IOP Publishing.

\bibitem[Ho et~al., 2017]{Ho2017}
Ho, M., Lall, U., Sun, X., and Cook, E. R. (2017).
\newblock Multiscale temporal variability and regional patterns in 555 years of conterminous U.S. streamflow. 
\newblock {\em Water Resources Research}, 53(4), 3047–3066.

\bibitem[Hornberger et~al., 2015]{Hornberger2015}
Hornberger, G.~M., Hess, D. J., and Gilligan, J.  (2024).
\newblock Water conservation and hydrological transitions in cities in the United States.
\newblock {\em Water Resources Research}, 51(6), 4635–4649.

\bibitem[Humphrey et~al., 2016]{Humphrey2016}
Humphrey, V., Gudmundsson, L., and Seneviratne, S.~I. (2016).
\newblock Assessing Global Water Storage Variability from GRACE: Trends, Seasonal Cycle, Subseasonal Anomalies and Extremes.
\newblock {\em Surveys in Geophysics}, 37, 357–395.

\bibitem[IDNR, 2007]{IDNR2007}
Indiana Department of Natural Resources [IDNR]. (2007).
\newblock Indiana Suggested Model Ordinance.
\newblock Accessed on 1/31/2025: https://www.in.gov/dnr/water /files/Model\_ordinance\_Final\_Draft\_7-2-07.doc

\bibitem[IDNR, 2015]{IDNR2015}
Indiana Department of Natural Resources [IDNR]. (2015).
\newblock Indiana’s Water Shortage Plan. 
\newblock Accessed on 1/31/2025: https://www.in.gov/dnr/water/files/watshplan.pdf 

\bibitem[IDNR, 2025]{IDNR2025}
Indiana Department of Natural Resources [IDNR]. (2025).
\newblock  Report on Indiana Water Use Efficiency and Conservation.
\newblock Accessed on 1/31/2025: https://www.in.gov/oucc/watersewer/ key-cases-by-utility/citizens-water-and-sewer-rates /\#Current\_Sewer\_Base\_Rates 

\bibitem[Indiana Office of Utility Consumer Counselor, 2025]{IndianaOffice2025}
Indiana Office of Utility Consumer Counselor. (2025).
\newblock  Citizens Water \& Sewer Rates.
\newblock Accessed on 1/31/2025: https://www.in.gov/dnr/water/lake-michigan/great-lakes-compact /report-on-indiana-water-use-efficiency-and-conservation/

\bibitem[Koppenjan and Groenewegen, 2005]{Koppenjan2005}
\newblock Koppenjan, J., and Groenewegen, J. (2005).
\newblock Institutional design for complex technological systems. 
\newblock {\em International Journal of Technology Policy and Management}, 5, 240.

\bibitem[Krueger et al., 2022]{Krueger2022}
\newblock Krueger, E.~H., McPhearson, T., and Levin, S.~A. (2022).
\newblock Integrated assessment of urban water supply security and resilience: towards a streamlined approach
\newblock {\em Environmental Research Letters}, 17(7), 075006.

\bibitem[Lawless et al., 2024]{Lawless2024}
\newblock Lawless, K.~L., Garcia, M., and White, D.~D. (2024).
\newblock Institutional analysis of water governance in the Colorado River Basin, 1922–2022.
\newblock {\em Frontiers in Water}, 6, 1451854.

\bibitem[Levin et~al., 2022]{Levin2022}
Levin, S.~A., Anderies, J.~M., Adger, N., Barrett, S., Bennett, E.~M.,
  Cardenas, J.~C., Carpenter, S.~R., Crépin, A.-S., Ehrlich, P., Fischer, J.,
  Folke, C., Kautsky, N., Kling, C., Nyborg, K., Polasky, S., Scheffer, M.,
  Segerson, K., Shogren, J., van~den Bergh, J., Walker, B., Weber, E.~U., and
  Wilen, J. (2022).
\newblock Governance in the {Face} of {Extreme} {Events}: {Lessons} from
  {Evolutionary} {Processes} for {Structuring} {Interventions}, and the {Need}
  to {Go} {Beyond}.
\newblock {\em Ecosystems}, 25(3):697--711.

\bibitem[Li et al., 2015]{Li2015}
Li, B., Rodell, M., and Famiglietti, J. S.   (2015).
\newblock Groundwater variability across temporal and spatial scales in the central and northeastern U.S.
\newblock {\em Journal of Hydrology}, 525, 769–780.

\bibitem[Magliocca et al., 2018]{Magliocca2018}
Magliocca, N.~R., Ellis, E.~C., Allington, G.~R.~H., de Bremond, A., Dell’Angelo, J., Mertz, O., Messerli, P., Meyfroidt, P., Seppelt, R., and Verburg, P.~H. (2018).
\newblock Closing global knowledge gaps: Producing generalized knowledge from case studies of social-ecological systems.
\newblock {\em Global Environmental Change}, 50, 1–14.

\bibitem[Marvel et al., 2023]{Marvel2023}
Marvel, K., Su, W., Delgado, R., Aarons, S., Chatterjee, A., Garcia, M., Hausfather, Z., Hayhoe, K., Hence, D., Jewett, E., Robel, A., Singh, D., Tripati, A., and Vose, R. (2023).
\newblock Chapter 2: Climate Trends
\newblock {\em Fifth National Climate Assessment.}.

\bibitem[Massmann, 2020]{Massmann2020}
Massmann, C. (2020).
\newblock Identification of factors influencing hydrologic model performance using a top‐down approach in a large number of U.S. catchments.
\newblock {\em Hydrological Processes}, 34, 4–20.

\bibitem[McCabe and Wolock, 2016]{McCabe2016}
McCabe, G.~J., and Wolock, D.~M. (2016).
\newblock Variability and Trends in Runoff Efficiency in the Conterminous United States. 
\newblock {\em Journal of the American Water Resources Association}, 52, 1046–1055.

\bibitem[McDonald et al., 2014]{McDonald2014}
McDonald, R.~I., Weber, K., Padowski, J., Flörke, M., Schneider, C., Green, P.~A., Gleeson, T., Eckman, S., Lehner, B., Balk, D., Boucher, T., Grill, G., and Montgomery, M. (2014).
\newblock Water on an urban planet: Urbanization and the reach of urban water infrastructure.
\newblock {\em Global Environmental Change}, 27, 96–105.

\bibitem[McNeely, 2023]{McNeely2023}
McNeely, E. (2023).
\newblock The History of Charlotte Water. 
\newblock Accessed on 2/12/2025: https://cltwaterblog.org/2023/04/the-history-of-charlotte-water/\#:\~:text=Charlotte\%20City\%20Council\%20purchased

\bibitem[Mesdaghi et al., 2022]{Mesdaghi2022}
Mesdaghi, B., Ghorbani, A., and de Bruijne, M. (2022).
\newblock Institutional dependencies in climate adaptation of transport infrastructures: an Institutional Network Analysis approach.
\newblock {\em Environmental Science and Policy}, 127, 120–136.

\bibitem[Miller, 1992]{Miller1992}
Miller, D. (1992).
\newblock The icarus paradox: How exceptional companies bring about their own downfall. 
\newblock {\em Business Horizons}, 35(1), 24–35.

\bibitem[Miller, 2017]{Miller2017}
Miller. (2017).
\newblock Water, sewer rates rising in Harrisburg area.
\newblock {\em PennLive}
\newblock Accessed on 3/28/2025: https://www.pennlive.com/news/2017/11/ water\_sewer\_rates\_rising\_in\_ha.html

\bibitem[Moore et al., 2024]{Moore2024}
Moore, M. L., Wang-Erlandsson, L., Bodin, Ö., Enqvist, J., Jaramillo, F., Jónás, K., Folke, C., Keys, P., Lade, S. J., Garcia, M. M., Martin, R., Matthews, N., Pranindita, A., Rocha, J. C., and Vora, S. (2024).
\newblock Moving from fit to fitness for governing water in the Anthropocene.
\newblock {\em Nature Water}, 2, 511-520.

\bibitem[Modarres, R, 2007]{Modarres2007}
Modarres, R. (2007).
\newblock Streamflow drought time series forecasting. 
\newblock {\em Stochastic Environmental Research and Risk Assessment}, 21: 223–233.

\bibitem[Municipal Water Leader, 2026]{MWL2026}
Municipal Water Leader. (2026).
\newblock The Central Role of the Israel Water Authority
\newblock https://municipalwaterleader.com/ the-central-role-of-the-israel-water-authority/

\bibitem[NOAA, 2012]{NOAA2012}
NOAA (2012).
\newblock Indianapolis, IN Weather Forecast Office, Hot and Dry Early Summer 2012. 
\newblock Accessed on 2/12/2025: https://www.weather.gov/ind/summer2012 

\bibitem[Novak Consulting Group, 2011]{Novak2011}
Novak Consulting Group (2011).
\newblock Municipal Financial Recovery Act Recovery Plan: City of Harrisburg. Prepared for the Commonwealth of Pennsylvania, Department of Community and Economic Development.
\newblock Accessed on 2/12/2025: https://media.pennlive.com/midstate \_impact/other/Act-47-Harrisburg.pdf

\bibitem[Oberlack et al., 2019]{Oberlack2019}
Oberlack, C., Sietz, D., Bonanomi, E. B., de Bremond, A., Dell’Angelo, J., Eisenack, K., Ellis, E. C., Epstein, G., Giger, M., Heinimann, A., Kimmich, C., Kok, M. T. J., Manuel-Navarrete, D., Messerli, P., Meyfroidt, P., Václavík, T., and Villamayor-Tomas, S. (2019).
\newblock Archetype analysis in sustainability research: meanings, motivations, and evidence-based policy making.
\newblock {\em Ecology and Society}, 24(26). 

\bibitem[Ostrom, 2009]{Ostrom2009}
Ostrom, E. (2009).
\newblock A general framework for analyzing sustainability of social-ecological systems. 
\newblock {\em Science}, 325, 419–422.

\bibitem[Ostrom, 1990]{Ostrom1990}
Ostrom, E. (1990).
\newblock {\em Governing the commons: The evolution of institutions for collective action.} Cambridge University Press.

\bibitem[Ostrom, 1995]{Ostrom1995}
Ostrom, E. (1995).
\newblock Designing Complexity to Govern Complexity. 
\newblock In S. Hanna and M. Munasinghe (Eds.), {\em Property Rights and the Environment: Social and Ecological Issues}, (Vol. 94). The World Bank.

\bibitem[Ostrom, 2007]{Ostrom2007}
Ostrom, E. (2007).
\newblock A diagnostic approach for going beyond panaceas. 
\newblock {\em Proceedings of the National Academy of Sciences of the United States of America}, 104(39), 15181–15187.

\bibitem[Ostrom, 2011]{Ostrom2011}
Ostrom, E. (2011).
\newblock Background on the Institutional Analysis and Development Framework. 
\newblock {\em Policy Studies Journal}, 39, 7–27.

\bibitem[Padowski and Jawitz, 2012]{Padowski2012}
Padowski, J. C., and Jawitz, J. W. (2012).
\newblock Water availability and vulnerability of 225 large cities in the United States
\newblock {\em Water Resources Research}, 48(12).

\bibitem[Peel et al., 2007]{Peel2007}
Peel, M. C., Finlayson, B. L., and McMahon, T. A.  (2007).
\newblock Updated world map of the Köppen-Geiger climate classification. 
\newblock {\em Hydrology and Earth System Sciences}, 11, 1633–1644.

\bibitem[Peixóto and Oort, 1984]{Peixóto1984}
Peixóto, J. P., and Oort, A. H. (1984).
\newblock Physics of climate.
\newblock {\em Reviews of Modern Physics}, 56, 365–429.

\bibitem[Pennsylvania Department of Environmental Protection, 2009]{PDEP2009}
Pennsylvania Department of Environmental Protection. (2009).
\newblock A drought history maps: 1980–present. 
\newblock Accessed on 1/31/2025: https://files.dep.state.pa.us/water/bsdw/ Drought/DroughtStatusMaps/PA\_Drought \_History\_Maps\_1980\_Present.pdf

\bibitem[Pennsylvania Department of Environmental Protection, 2025]{PDEP2025}
Pennsylvania Department of Environmental Protection. (2025).
\newblock Pennsylvania State Water Plan Principles 
\newblock Accessed on 3/28/2025: https://spcwater.org/wp-content/uploads/2020/01/StateWtrPlan.pdf

\bibitem[Peterson and Muzzini, 2005]{Peterson2005}
Peterson, G., and Muzzini, E. (2005).
\newblock Decentralizing basic infrastructure services.
\newblock {\em East Asia Decentralizes: Making Local Government Work}, (pp. 209–236). The World Bank.

\bibitem[QC Water, 2025]{QCWater2025}
QC Water. (2025).
\newblock Renew Brew.  
\newblock Accessed on 1/31/2025: https://www.drinkqcwater.org/ 

\bibitem[Quan et al., 2013]{Quan2013}
Quan, C., Han, S., Utescher, T., Zhang, C., and Liu, Y.-S. (2013).
\newblock Validation of temperature–precipitation based aridity index: Paleoclimatic implications.
\newblock {\em Palaeogeography, Palaeoclimatology, Palaeoecology}, 386, 86–95.

\bibitem[Raftelis, 2019]{Raftelis2019}
Raftelis. (2019).
\newblock Stormwater Fee Proposal and Implementation Plan
\newblock Accessed on 3/28/2025: https://capitalregionwater.com/wp-content/uploads/2019/06/2019-06-19-Stormwater-Fee-Proposal-and-Implementation-Plan.pdf

\bibitem[Raftelis, 2025]{Raftelis2025}
Raftelis. (2025).
\newblock Capital Region Water.
\newblock Accessed on 3/28/2025: https://www.raftelis.com/work/capital-region-water/

\bibitem[Rice and Emanuel, 2017]{Rice2017}
Rice, J. S., and Emanuel, R. E.  (2017).
\newblock How are streamflow responses to the E l N ino S outhern O scillation affected by watershed characteristics? 
\newblock {\em Water Resources Research}, 53, 4393–4406.

\bibitem[Rice et al., 2016]{Rice2016}
Rice, J. S., Emanuel, R. E., and Vose, J. M. (2016).
\newblock The influence of watershed characteristics on spatial patterns of trends in annual scale streamflow variability in the continental U.S. 
\newblock {\em Journal of Hydrology}, 540, 850–860.

\bibitem[Rockström et al., 2014]{Rockström2014}
Rockström, J., Falkenmark, M., Allan, T., Folke, C., Gordon, L., Jägerskog, A., Kummu, M., Lannerstad, M., Meybeck, M., Molden, D., Postel, S., Savenije, H. H. G., Svedin, U., Turton, A., and Varis, O.  (2014).
\newblock The unfolding water drama in the Anthropocene: towards a resilience‐based perspective on water for global sustainability. 
\newblock {\em Ecohydrology}, 7, 1249–1261.

\bibitem[Rodriguez et al., 2011]{Rodriguez2011}
Rodriguez, A. A., Cifdaloz, O., Anderies, J. M., Janssen, M. A., and Dickeson, J. (2011).
\newblock Confronting management challenges in highly uncertain natural resource systems: A robustness–vulnerability trade-off approach.
\newblock {\em Environmental Modeling and Assessment}, 16(1), 15–36.

\bibitem[Rudnick et al., 2025]{Rudnick2025}
Rudnick, J., Heikkila, T., Koebele, E., Morrison, T., and Batavia, C.  (2025).
\newblock Assessing the State and Efficacy of Climate Governance Research and Practice in the Sacramento-San Joaquin Delta.
\newblock {\em San Francisco Estuary and Watershed Science}, 23.

\bibitem[Saki et al., 2023]{Saki2023}
Saki, S. A., Sofia, G., and Anagnostou, E. N. (2023).
\newblock Characterizing CONUS-wide spatio-temporal changes in daily precipitation, flow, and variability of extremes.
\newblock {\em Journal of Hydrology}, 626, 130336.

\bibitem[Sietz et al., 2019]{Sietz2019}
Sietz, D., Frey, U., Roggero, M., Gong, Y., Magliocca, N., Tan, R., Janssen, P., and Václavík, T.  (2019).
\newblock Archetype analysis in sustainability research: methodological portfolio and analytical frontiers. 
\newblock {\em Ecology and Society}, 24(34). 

\bibitem[Singh et al., 2021]{Singh2021}
Singh, S., Abebe, A., Srivastava, P., and Chaubey, I. (2021).
\newblock Effect of ENSO modulation by decadal and multi-decadal climatic oscillations on contiguous United States streamflows.
\newblock {\em Journal of Hydrology. Regional Studies, }, 36, 100876.

\bibitem[Steffen et al., 2015]{Steffen2015}
Steffen, W., Broadgate, W., Deutsch, L., Gaffney, O., and Ludwig, C. (2015).
\newblock The trajectory of the Anthropocene: The Great Acceleration.
\newblock {\em The Anthropocene Review}, 2, 81–98.

\bibitem[Swain et al., 2025]{Swain2025}
Swain, D. L., Prein, A. F., Abatzoglou, J. T., Albano, C. M., Brunner, M., Diffenbaugh, N. S., Singh, D., Skinner, C. B., and Touma, D. (2025).
\newblock Hydroclimate volatility on a warming Earth.
\newblock {\em Nature Reviews. Earth and Environment}, 6, 35–50.

\bibitem[Treuer et al., 2017]{Treuer2017}
Treuer, G., Koebele, E., Deslatte, A., Ernst, K., Garcia, M., and Manago, K. (2017).
\newblock A narrative method for analyzing transitions in urban water management: The case of the Miami-Dade Water and Sewer Department.  
\newblock {\em Water Resources Research}, 53, 891–908.

\bibitem[Turlington et al., 2017]{Turlington2017}
Turlington, M. W., de Neufville, R., and Garcia, M. (2017).
\newblock Flexible Design of Water Infrastructure Systems. 
\newblock In S. Islam and K. Madani (Eds.), {\em Contingent Complexity and Prospects for Water Diplomacy: Understanding and Managing Risks and Opportunities for an Uncertain Water Future}, Anthem Press.

\bibitem[U.S. vs Capital Region Water, 2015]{US2015}
United States \& Commonwealth of Pennsylvania Department of Environmental Protection v. City of Harrisburg \& Capital Region Water. (2015).
\newblock No. 1:15-cv-00291 (M.D. Pa. 2015). 
\newblock Accessed on 1/31/2025: https://www.epa.gov/sites /default/files/2015-02/documents/cityofharrisburg-cd.pdf

\bibitem[Unterreiner, 2011]{Unterreiner2011}
Unterreiner, J. (2011).
\newblock Indiana's Water Shortage Plan. 2011 Symposium on Data-Driven Approaches to Droughts. Paper 9.
\newblock Accessed on 1/31/2025: http://docs.lib.purdue.edu/ddad2011/9

\bibitem[USDA, 1998]{USDA1988}
USDA (1988).
\newblock Indiana Agricultural Statistics - 1988. 
\newblock Accessed on 1/31/2025: https://www.nass.usda.gov/Statistics\_by\_State/Indiana/ Publications/Annual\_Statistical\_Bulletin/hist\_pdf /as\_1988\_ebook.pdf

\bibitem[USGS, 1992]{USGS1992}
U.S. Geological Survey [USGS]. (1992).
\newblock Description and Effects of 1988 Drought on Ground-water Levels, Streamflow and Reservoir Levels in Indiana.
\newblock Accessed on 1/31/2025: https://pubs.usgs.gov/wri/wri914100/pdf/wrir\_91-4100\_a.pdf 

\bibitem[USGS, 2005]{USGS2005}
U.S. Geological Survey [USGS]. (2005).
\newblock The Drought of 1998–2002 in North Carolina — Precipitation and Hydrologic Conditions. 
\newblock Accessed on 1/31/2025: https://pubs.usgs.gov/sir/2005/5053/pdf/SIR2005-5053.pdf

\bibitem[Veissière et al., 2020]{Veissière2020}
Veissière, S. P. L., Constant, A., Ramstead, M. J. D., Friston, K. J., and Kirmayer, L. J.  (2020).
\newblock Thinking through other minds: A variational approach to cognition and culture.
\newblock {\em The Behavioral and Brain Sciences}, 43, e90.

\bibitem[Votteler, 2023]{Votteler2023}
Votteler, T. H. (2023).
\newblock Hands across the Water: How the 57-Year Dispute over the Edwards Aquifer Began, Persisted, and Was Resolved.
\newblock {\em Water}, 15, 1835.

\bibitem[Water Research Foundation, 2013]{WRF2013}
Water Research Foundation. (2013).
\newblock Defining and Enhancing the Safe Yield of a Multi-Use, Multi-Reservoir Water Supply. 
\newblock Accessed on 1/31/2025: https://www.waterrf.org/research/projects/defining-and-enhancing-safe-yield-multi-use-multi-reservoir-water-supply

\bibitem[WBTV 3, 2011]{WBTV2011}
WBTV 3. (2011).
\newblock Charlotte- Mecklenburg Utilities Launches Irrigation Incentive program. 
\newblock Accessed on 1/31/2025: https://www.wbtv.com/story /15361888/charlotte-mecklenburg-utilities-launches-irrigation-incentive-program/

\bibitem[Wiechman et~al., 2024]{Wiechman2024}
Wiechman, A., Alonso~Vicario, S., Anderies, J.~M., Garcia, M., Azizi, K., and Hornberger, G. (2024).
\newblock Institutional {Dynamics} {Impact} the {Response} of {Urban}
  {Socio}-{Hydrologic} {Systems} to {Supply} {Challenges}.
\newblock {\em Water Resources Research}, 60(2):e2023WR035565.

\bibitem[Wiechman et~al., 2024b]{WiechmanAlonso2024}
Wiechman, A., Vicario, S. A., and Koebele, E. A. (2024).
\newblock The Role of Intermediate Collaborative Forums in Polycentric Environmental Governance. 
\newblock {\em Journal of Public Administration Research and Theory}, 34, 196–210.

\bibitem[Williams et~al., 2021]{Williams2021}
Williams, J. H., Jones, R. A., Haley, B., Kwok, G., Hargreaves, J., Farbes, J., and Torn, M. S. (2021).
\newblock Carbon‐Neutral Pathways for the United States.
\newblock {\em AGU Advances}, 2.

\bibitem[Yao et~al., 2020]{Yao2020}
Yao, L., Libera, D. A., Kheimi, M., Sankarasubramanian, A., and Wang, D. (2020).
\newblock The Roles of Climate Forcing and Its Variability on Streamflow at Daily, Monthly, Annual, and Long‐Term Scales.
\newblock {\em Water Resources Research}, 56(7):e2020WR027111.

\bibitem[Young, 2002]{Young2002}
Young, O. R. (2002).
\newblock {\em The institutional dimensions of environmental change: fit, interplay, and scale.} MIT Press.


\end{thebibliography}
\end{document}